\documentclass[12pt]{iopart}

\usepackage{graphicx}% Include figure files
\usepackage{dcolumn}% Align table columns on decimal point
\usepackage{bm}% bold math
\usepackage{color}
\usepackage{epstopdf}
\usepackage{amsfonts}
\usepackage{amssymb}
\usepackage{mathrsfs}
\usepackage{amsthm}
\usepackage{subfigure}

\usepackage[caption=false,subrefformat=parens,labelformat=parens]{subfig}
\usepackage{float}
\usepackage{upgreek}

\usepackage{enumerate}

\begin{document}

\title{Estimating Entropy Production Rates   with  First-Passage Processes}

% TODO: write the author list here. Use initials + surname format.
% Separate subsequent authors by a comma, omit comma at the end of the list.
% Mark the corresponding author with a superscript *.
\author{Izaak Neri}

\address{Department of Mathematics, King’s College London, Strand, London, WC2R 2LS, UK}

\ead{izaak.neri@kcl.ac.uk}

% For convenience during refereeing: line numbers
%\linenumbers

\begin{abstract}     We consider the problem of estimating the mean entropy production rate in a nonequilibrium process from the measurements of first-passage quantities associated with a single current.      For first-passage processes with    large thresholds,  Refs.~\cite{roldan2015decision, neri2021universal} identified  a ratio of first-passage observables --- involving the mean first-passage time, the splitting probability, and the  first-passage thresholds--- that lower bounds the entropy production rate and is an unbiased estimator  of the entropy production rate when applied to a current that is proportional to the stochastic entropy production.   Here, we show that also at finite thresholds, a finite number of realisations of the nonequilibrium process, and for currents that are not proportional to the stochastic entropy production,  first-passage ratios can accurately estimate the rate of dissipation.  In particular,  first-passage ratios capture a finite fraction of the total entropy production rate   in regimes far from thermal equilibrium where  thermodynamic uncertainty ratios    capture a negligible fraction of the total entropy production rate.  Moreover, we show that first-passage ratios incorporate nonMarkovian statistics in the estimated value of the dissipation rate, which are difficult to include in estimates based on    Kullback-Leibler divergences.    Taken together, we  show that entropy production  estimation with first-passage ratios is complementary to  estimation methods based on thermodynamic uncertainty ratios and Kullback-Leibler divergences.    
\end{abstract}

\section{Introduction}      
Empirical measurements of   dissipated heat in  nonequilibrium, mesoscopic systems are challenging.       The obstacle is that   heat, defined as the remainder energy  required to satisfy the first law of thermodynamics  \cite{callen1985thermodynamics}, is  microscopically the energy exchanged between a system and its environment  through  interactions with a     large number of  microscopic  degrees of freedom \cite{sekimoto2010stochastic}.   Therefore,   contrarily to the work performed on a system, heat   cannot be measured directly.

To  illustrate the empirical difficulty of measuring  heat in small systems, let us consider the example of a molecular motor in a living cell.   A molecular motor    dissipates at a power of  $10^{-18} W$ \cite{hwang2017quantifying, hwang2018energetic, hwang2019correction}, which is several orders of magnitude smaller than the $10^{-6}W$ sensitivity   of calorimeters~\cite{zogg2004isothermal, maskow2015does}.      Nevertheless, even if it were possible to  measure heat  at an accuracy of   $10^{-18} W$, then  still it would not be possible to measure the rate of dissipation of a molecular motor in a living cell, as  in living cells a large number of dissipative processes take place at once.      Therefore, it is important to develop  alternative methods to measure dissipation in nonequilibrium, mesoscopic systems. 

Instead of measuring the rate $\dot{s}$ of dissipation, also known as the entropy production rate, with calorimeters, the entropy production rate can be estimated from  the  trajectories of slow, coarse-grained degrees of freedom, which is an example of a   thermodynamic inference problem~\cite{seifert2019stochastic}.   In this paper, we consider  entropy production estimation   based on the trajectories of a  {\it single} current $J$ in a  stationary  process.    An observer measures, e.g.,  the current in an electrical wire or the position of a molecular motor bound to a biofilament, and aims to estimate the rate of dissipation from the obtained trajectories.   If the average current $\langle J(t)\rangle\neq 0$, then  the system is driven out of equilibrium.  However, in general, it is not possible to estimate the rate of dissipation from the  average current, except when the current's thermodynamic affinity is known and the multiple currents of a system are uncoupled.   On the other hand, when the whole trajectory of the current  $J$ is available, then  it is possible to estimate the entropy production rate from   fluctuations in  the current~\cite{gingrich2017inferring}.    

An estimator of dissipation is a real-valued functional  evaluated on the trajectories of a nonequilibrium process.   In the present paper we limit ourselves to estimators that are evaluated on the trajectories of    a  single current $J$, and  we also consider estimators that do not require knowledge about  the model that determines the statistics of the nonequilibrium process $X$.      These two limitations model physical scenarios for which only partial information about the nonequilibrium process is available, such as,  the case when the position of a molecular motor is measured, but  the chemical states of its motor heads are not known.

The quality of an estimator is determined by its bias, i.e., the distance between the average value of the estimator and the rate of dissipation.      Since we consider problems with partial information,  it is in general not possible to obtain an unbiased estimator.   In fact, when observing a single current,  a good estimator of dissipation will underestimate the amount of dissipation in the process, as  in general there exist other dissipative processes that do not directly influence   the measured current.          A notable exception is when the current $J$ is proportional to the stochastic entropy production.   In this case, we expect a good estimator to be unbiased.    Lastly, let us mention that the bias of an estimator can depend on the number of trajectories available.   In other words, an estimator may have a small bias in the limit of an infinite number of realisations of a nonequilibrium process, while in situations of finite data the same estimator may have a large bias.

As the entropy production estimation  problem  has been considered before, we first review two    estimators that have been well studied  in the   literature.   Subsequently,   we present an estimator based on first-passage processes that has been  less studied, and which  constitutes the main topic of this paper.

The first estimator that we review is  the  {\it Kullback-Leibler divergence} between the probability to observe a trajectory of a current and the probability to observe its time-reversed trajectory, see e.g.,~Refs.~\cite{gomez2008lower, PhysRevLett.105.150607, martinez2019inferring, roldan2021quantifying, ehrich2021tightest, harunari2022learn}.    In general, it is however impractical to evaluate the Kullback-Leibler divergence.  This is because the statistics of a single current $J$ are nonMarkovian, which complicates significantly the evaluation of the probability to observe a certain trajectory.      Because of this complication, nonMarkovian effects in the current are often neglected by evaluating \cite{gomez2008lower,PhysRevLett.105.150607, martinez2019inferring}
 \begin{equation}
 \hat{s}_{\rm KL} := \sum_{j\in \mathcal{J}}\dot{n}_j\log \frac{\dot{n}_j}{\dot{n}_{-j}}, \label{eq:KL}
 \end{equation}  
 where $\dot{n}_j$ is the rate at which the current  $J$ makes a  jump of size $j$, and $\mathcal{J}$ is the set of  possible jump sizes.    The  estimator  (\ref{eq:KL})  neglects nonMarkovian statistics in $J$'s trajectory, and consequently the estimator is biased, i.e., $\hat{s}_{\rm KL}\leq \dot{s}$.

A second estimator of dissipation that we review is the {\it Thermodynamic Uncertainty Ratio}~(TUR)  \cite{pietzonka2016universal, gingrich2017inferring, seifert2019stochastic, van2020entropy,  manikandan2020inferring}
 \begin{equation}
 \hat{s}_{\rm TUR} := \frac{2\overline{j}^2t}{\sigma^2_{J(t)}}, \label{eq:TUR}
 \end{equation}  
 where $J(t)$ is a stochastic current at time $t$, $\overline{j} = \langle J(t)\rangle/t$ is the average current rate, and $\sigma^2_{J(t)} = \langle J^2(t)\rangle - \langle J(t)\rangle^2$ is the variance of the current.     
   The TUR has been   proposed as an estimator of dissipation as it lower bounds the entropy production rate  in stationary Markov jump processes and  overdamped Langevin processes, i.e., $\hat{s}_{\rm TUR}\leq \dot{s}$.      However,  the bias of the TUR  is large for systems far from thermal equilibrium.     For example,  it was shown that  $ \hat{s}_{\rm TUR}$ captures  20 \% of the total dissipation of molecular motors, such as kinesin and myosin~\cite{hwang2018energetic, hwang2019correction}, and the bias of the TUR remains large when  $J$ equals the stochastic entropy production~\cite{pigolotti2017generic}.

In an attempt at resolving the  issues in the bias of the Kullback-Leibler divergence and the TUR,  we study in this paper 
the  {\it First-Passage Ratio}~(FPR)~\cite{roldan2015decision, neri2021universal}
\begin{equation}
\hat{s}_{\rm FPR}(\ell_-,\ell_+) :=  \left\{\begin{array}{ccc}\frac{\ell_+}{\ell_- }\frac{1}{\langle T_J\rangle} \left| \ln p_-\right|, && \overline{j}>0, \\  \frac{\ell_-}{\ell_+ }\frac{1}{\langle T_J\rangle} \left| \ln p_+\right|,&& \overline{j}<0, \end{array}\right.  \label{eq:FPR}
\end{equation}
where $T_J$ is the first time when the 
stochastic current $J$ leaves the open interval $(-\ell_-,\ell_+)$ with $\ell_-,\ell_+>0$,  $p_-$  is the probability that $J$ first goes below the negative threshold $-\ell_-$ before exceeding the positive value $\ell_+$, and $p_+=1-p_-$.     For  stationary Markov jump processes and  in the limit of large thresholds $\ell_-$ and $\ell_+$,  the FPR bounds the dissipation rate from below, i.e.,    $\hat{s}_{\rm FPR}\leq \dot{s}$~\cite{roldan2015decision, neri2021universal}.   In addition,   the FPR is an unbiased estimator when $J$ is proportional to the stochastic entropy production \cite{roldan2015decision, neri2021universal}, i.e., $\hat{s}_{\rm FPR} = \dot{s}$, a result that also holds in the limit of large thresholds.     

\begin{figure*}[t!]\centering
  \subfigure[Random walk on a two-dimensional lattice]{  \includegraphics[width=0.5\textwidth]{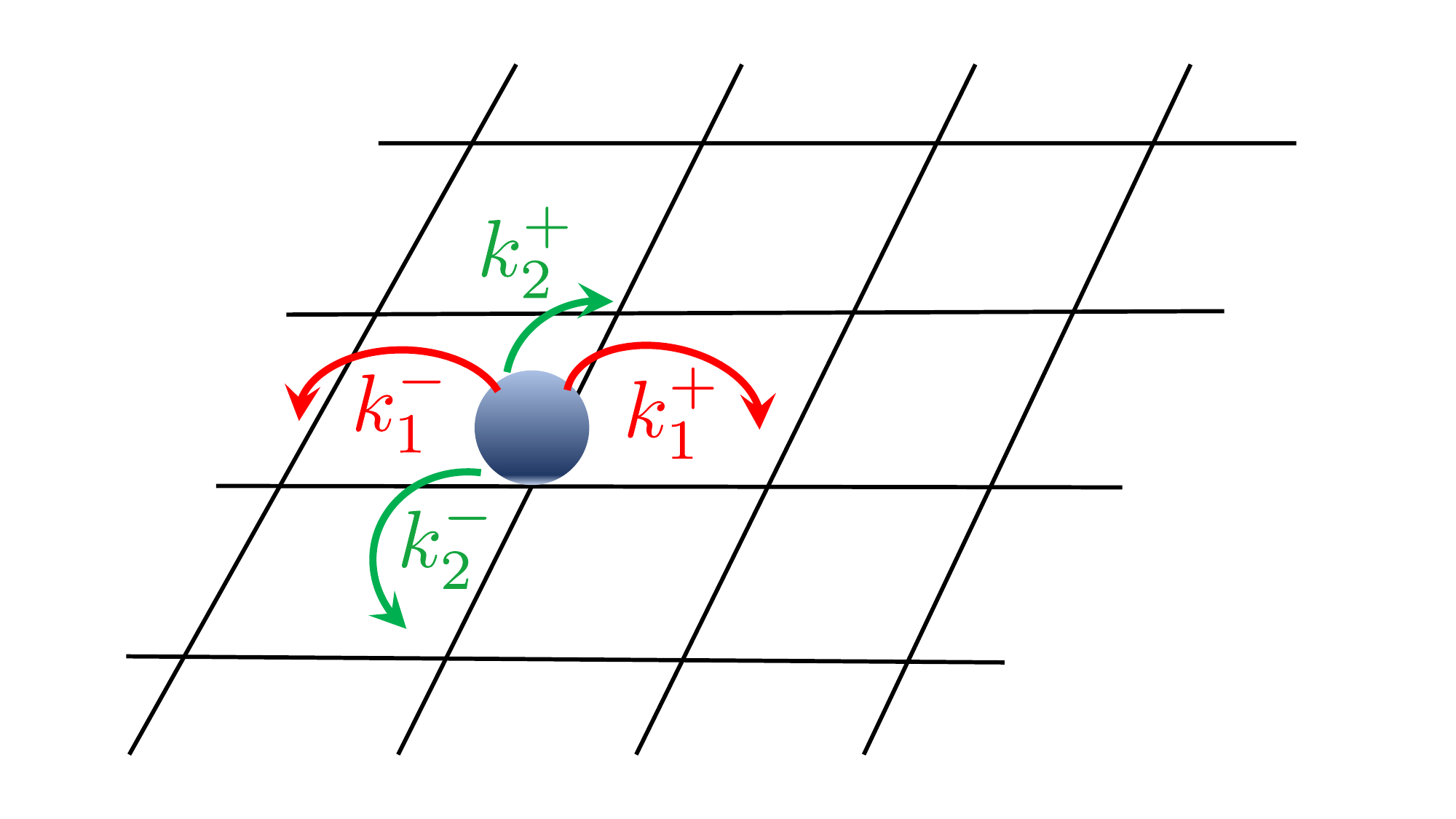}}
    \subfigure[Six state model for molecular motors with two motor heads, such as, conventional kinesin]{  \includegraphics[width=0.5\textwidth]{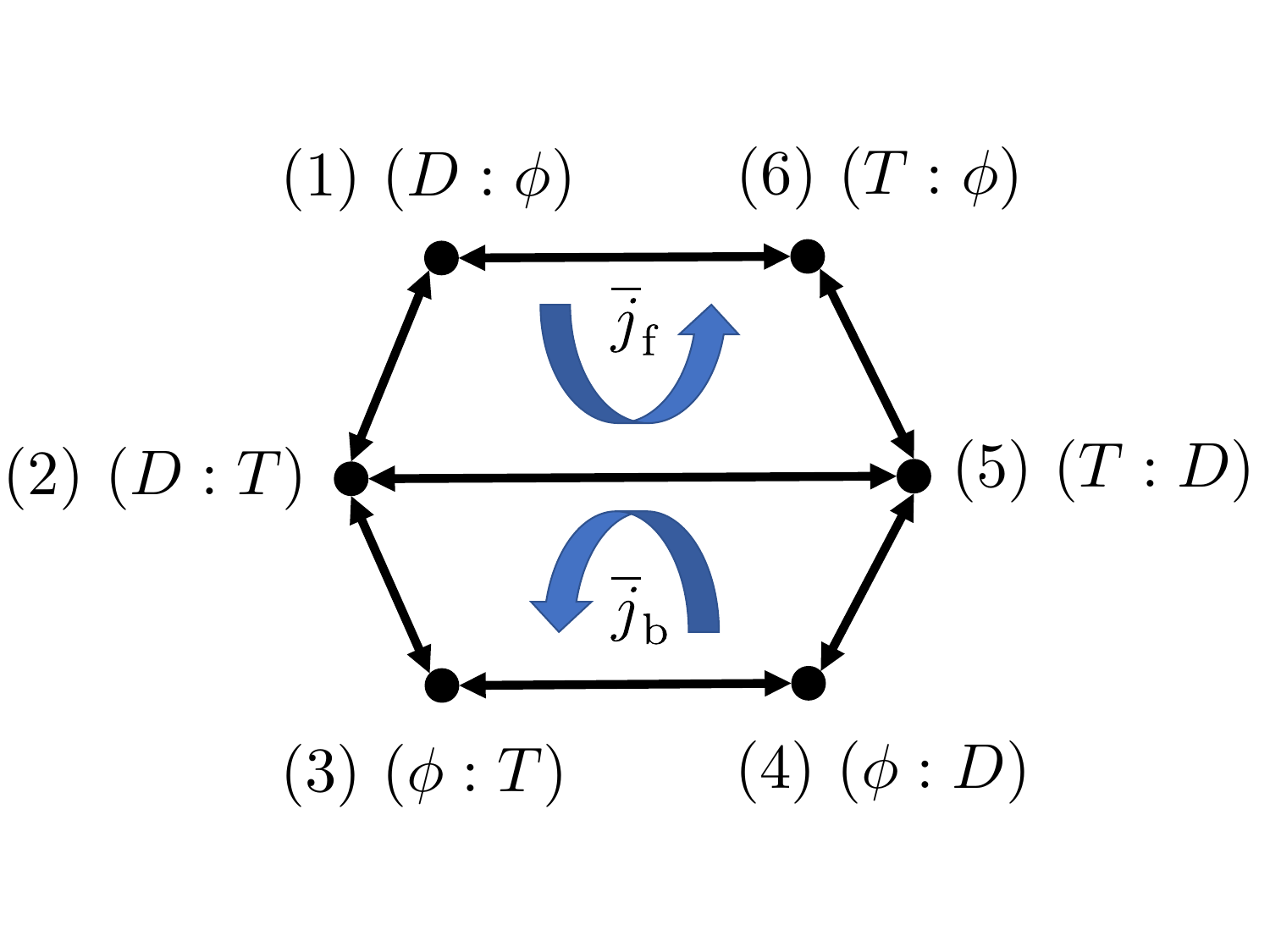}}
\caption{Illustration of the two paradigmatic examples of nonequilibrium processes that we consider in this paper.  Subfigure (a): Random walker jumping on a two-dimensional lattice at rates $k^{+}_1$, $k^-_1$, $k^+_2$, and $k^{-}_2$.    Subfigure (b):   Markov jump process with six states, representing the different chemical configurations of the two motor heads of a molecular motor bound to a biofilament \cite{PhysRevLett.98.258102, hwang2018energetic}.   The different motor states are  ATP bound (T), ADP bound (D), and nucleotide free ($\phi$).    Since there are six states, we also label them with $\left\{1,2,3,4,5,6\right\}$ as indicated in the figure.  The motor moves forwards along the cycle  ($2\rightarrow 5\rightarrow 6\rightarrow 1\rightarrow 2$)  and moves backwards along the cycle ($5\rightarrow 2\rightarrow 3\rightarrow 4\rightarrow 5$) at rates $\overline{j}_{\rm f}$ and $\overline{j}_{\rm b}$, respectively. } \label{fig:examples}
\end{figure*}

Although it was established that $\hat{s}_{\rm FPR}$ equals $\dot{s}$ for large thresholds and for currents that are proportional to the stochastic entropy production, this is not sufficient to conclude that  $\hat{s}_{\rm FPR}$ is a useful estimator of dissipation.   Therefore, in the  present paper, we determine  the bias in the estimator  $\hat{s}_{\rm FPR}$ for: (i) generic currents that are not proportional to the entropy production; (ii) 
 finite thresholds $\ell_+$ and $\ell_-$; and (iii) a finite number of realisations of the nonequilibrium process.    To this aim, we estimate dissipation based on the FPR in 
two   paradigmatic examples of   nonequilibrium Markov jump processes, namely, a random walker on a two-dimensional  lattice  [see Fig.~\ref{fig:examples}(a)] and a model for a  molecular motor with two motor heads  [see Fig.~\ref{fig:examples}(b)].   To evaluate the quality of $\hat{s}_{\rm FPR}$, we compare the bias in the estimates of dissipation based on the  FPR with those based on the TUR and the Kullback-Leibler divergence.

The paper is organised as follows:  In Sec.~\ref{sec:2}, we introduce the setup of stationary Markov jump processes that we use in this paper, and we define entropy production in this setup.     In Sec.~\ref{sec:3}, we define  estimators of entropy production based on the FPR, the TUR, and the Kullback-Leibler divergence.  In Sec.~\ref{sec:4}, we make a study of  $\hat{s}_{\rm FPR}$  as an estimator of the entropy production rate in a two-dimensional random walk process, and in Sec.~\ref{sec:5}, we make an analogous study for a model of a molecular motor with two motor heads, such as, conventional kinesin.    We end the paper with a discussion in Sec.~\ref{sec:6}, where we compare the advantages and disadvantages of the different estimators for dissipation.       The paper also contains  three appendices with detailed information on some of the calculations and the parameters used in the models.

\section{Entropy production in stationary Markov jump processes} \label{sec:2}

We first present in Sec.~\ref{sec:21}  the mathematical setup on which we work in this paper, and in Sec.~\ref{sec:22} we  discuss the physical context of the mathematical setup.

\subsection{General system setup}\label{sec:21} 
We focus in this paper on stationary Markov jump processes \cite{bremaud2013markov}, which  play an important role in  stochastic thermodynamics \cite{seifert2012stochastic, peliti2021stochastic}; doing so, we neglect  the possibility of a Hamiltonian  evolution~\cite{PhysRev.99.578}.

Let $X(t)$ be a Markov jump process taking values in a discrete set $\mathcal{X}$ \cite{bremaud2013markov}, with $t\geq 0$ a time index.   We denote by $\mathbb{P}$ the probability measure that generates the statistics of $X$, and averages with respect to $\mathbb{P}$ are denoted by $\langle \cdot \rangle_{\mathbb{P}}$, or more briefly  $\langle \cdot \rangle$.   Trajectories of $X(t)$ are denoted by $X^t_0$.

We assume that the process $X$ is regular, i.e., it has a finite number of jumps within each  time interval of finite length.   The  probability distribution $p_{X(t)}(x) = p(x;t)$  of $X(t)$ solves  Kolmogorov's  differential system \cite{bremaud2013markov}
\begin{equation}
\partial_t p(x;t) = \sum_{y \in \mathcal{X}\setminus x}p(y;t) w_{y\rightarrow x} - p(x;t) \sum_{y \in \mathcal{X}\setminus x}w_{x\rightarrow y}, \quad {\rm for} \quad x\in\mathcal{X}, \label{eq:stat}
\end{equation} 
where $w_{x\rightarrow y}\in \mathbb{R}^{+}$ are the transition rates to jump from  $x$ to $y$.  We consider time-homogeneous Markov jump processes for which the rates $w_{y\rightarrow x}$ are independent of time~$t$.  

 We assume that $X$ admits  a unique stationary probability distribution $p_{\rm ss}(x)$ with $p_{\rm ss}(x)>0$ for all $x\in\mathcal{X}$,  and we set $p_{X(0)}(x) = p_{\rm ss}(x)$, so that the process is stationary.     As shown in Ref.~\cite{bremaud2013markov}, in this setup it holds that for  generic initial conditions the stationary state  is reached in a finite time, i.e.,~$\lim_{t\rightarrow \infty}p(x;t) = p_{\rm ss}(x)$, and moreover the ergodic theorem implies that $\lim_{t\rightarrow \infty}t^{-1}\int^t_0f(X(s)){\rm d}s = \sum_{x\in\mathcal{X}}f(x)p_{\rm ss}(x) $ for all initial distributions and real-valued functions $f$ with $\sum_{x\in\mathcal{X}}|f(x)|p_{\rm ss}(x)<\infty$~\cite{bremaud2013markov}.  
  Note that for finite sets $\mathcal{X}$, a sufficient requirement for the uniqueness of  the stationary probability distribution   is  that  the graph of nonzero transitions $w_{x\rightarrow y}$ is strongly connected --- a property that is often called irreducibility--- and  for countable sets $\mathcal{X}$ we also require that the process is  $t$-positive recurrent, i.e., the mean  time for $X$  to return to its initial state is finite~\cite{bremaud2013markov}.

The last assumption we require is that the  Markov jump process is reversible in the sense that \begin{equation}
w_{x\rightarrow y} = 0 \Leftrightarrow w_{y\rightarrow x} = 0, \quad  \forall x,y\in\mathcal{X}. \label{eq:reversible}
\end{equation}

 Given the above assumptions, we can define the  stochastic entropy production by~\cite{schnakenberg1976network, peliti2021stochastic}
\begin{equation}
S(t) := \frac{1}{2} \sum_{x,y\in \mathcal{X}}  J_{x\rightarrow y}(t)\ln \frac{p_{\rm ss}(x)w_{x\rightarrow y}}{p_{\rm ss}(y)w_{y\rightarrow x}}  ,\label{eq:Stot}
\end{equation} 
where  the edge currents 
\begin{equation}
J_{x\rightarrow y}(t) = N_{x\rightarrow y}(t) - N_{y\rightarrow x}(t) \label{eq:edge}
\end{equation}
denote the difference between the number of times $N_{x\rightarrow y}(t)$  the process $X$ has jumped from $x$ to $y$ in the time interval $[0,t]$ and the number of times $N_{y\rightarrow x}(t)$ the process $X$ has jumped from $y$ to $x$ in the same time interval.      We denote the mean entropy production rate by 
\begin{equation}
\dot{s} := \langle S(t)\rangle_{\mathbb{P}}/t = \sum_{x,y\in \mathcal{X}} p_{\rm ss}(x)w_{x\rightarrow y} \ln \frac{p_{\rm ss}(x)w_{x\rightarrow y}}{p_{\rm ss}(y)w_{y\rightarrow x}} , \label{eq:sdotx}
\end{equation}
which will be the main quantity of interest in this paper.

\subsection{Physical interpretation  of $\dot{s}$ based on local detailed balance}\label{sec:22}
To give physical content to the  entropy production  rate $\dot{s}$---  in the  sense of dissipation or the entropy production  of the second law of thermodynamics ---  it is necessary to make physical assumptions about the nature of the environment and the interactions between the system and its environment.     It should be emphasised that these physical assumptions are mathematically superfluous, as they do not alter the model setup given by Eq.~(\ref{eq:stat}).  Indeed,  Eq.~(\ref{eq:stat}) provides a coarse-grained description of a system for which environmental degrees of freedom have already been integrated out.    Nevertheless, to understand the physical relevance of $\dot{s}$, it is  necessary to consider how  Eq.~(\ref{eq:sdotx})   appears in the broader picture of a system coupled to an  environment.  

  We assume that $X\in \mathcal{X}$ represents the slow, coarse-grained degrees of freedom of a mesoscopic system that is in contact with one or more equilibrium reservoirs.     A given mesoscopic configuration $x\in\mathcal{X}$ has an energy $u(x)$ and an  entropy $s_{\rm int}(x)$ due to the existence of fast, internal degrees of freedom.   
  
 Following Refs.~\cite{maes2003time, seifert2012stochastic, maes2020local, peliti2021stochastic},  we use the principle of local detailed balance, which states that 
\begin{equation}
\frac{w_{x\rightarrow y}}{w_{y \rightarrow x}} = e^{s_{xy}}, \quad  {\rm with}\quad x,y\in\mathcal{X},\label{eq:phys}
\end{equation}
where  $s_{xy} (=-s_{yx})$ is the change of entropy in  the environment and  the fast, internal degrees of freedom of the system   when $X$ jumps from $x$ to $y$.     Note that Eq.~(\ref{eq:phys}) does not require that the system is near equilibrium --- as $\dot{s}$ can be large ---  but it does require that the environment is equilibrated.   To have a system far from thermal equilibrium that interacts with an environment in equilibrium, we require,  in general,  weak coupling between system and environment and fast relaxation of the degrees of freedom in the environment. 

  The precise form of $s_{xy}$ depends on the physical problem of interest.    For a system in thermal equilibrium, 
\begin{equation}
s_{xy} = \frac{f(x) - f(y)}{\mathsf{T}_{\rm env}}, \label{eq:sxy}
\end{equation}
where $f(x) = u(x) - \mathsf{T}_{\rm env} \: s_{\rm int}(x)$ is the free energy of the $x$-th state, and $\mathsf{T}_{\rm env}$ is the temperature of the environment.   Note that we use natural units for which Boltzmann's constant is set to one.     For a nonequilibrium system in contact with a thermal reservoir, Eq.~(\ref{eq:sxy}) extends to 
\begin{equation}
s_{xy} =s_{\rm int}(y)-s_{\rm int}(x) -\frac{Q}{\mathsf{T}_{\rm env}}, \label{eq:sxy2}
\end{equation}
where  $Q$ is the net heat absorbed by the system from the environment.   Equation~(\ref{eq:sxy2}) provides a direct link between the entropy production rate $\dot{s}$, as defined in Eq.~(\ref{eq:sdotx}), and the rate of  heat dissipated to the environment, 
\begin{equation}
\dot{s} = -\frac{\langle Q(t)\rangle}{t \mathsf{T}_{\rm env}}.
\end{equation} 
Therefore, in this paper, we call $\dot{s}$ the entropy production rate, as well as, the rate of dissipation.

As a last example, let us consider the case of a molecular motor that interacts with $m$ chemostats at chemical potentials $\mu_{\alpha}$, with $\alpha=1,2,\ldots,m$, and an external agent that exerts a mechanical force $f_{\rm mech}$ on the motor.  In this example \cite{pietzonka2016universal}
\begin{equation}
s_{xy} =  \frac{f(x) - f(y)}{\mathsf{T}_{\rm env}}  + \sum^m_{\alpha=1} \frac{\mu_{\alpha}}{\mathsf{T}_{\rm env}}n_\alpha(x,y) - \frac{f_{\rm mech} \delta(x,y)}{\mathsf{T}_{\rm env}} ,  \label{eq:physa}
\end{equation}
where $n_\alpha(x,y) $ is the number of particles exchanged with the $\alpha$-th reservoir when $X$ jumps from $x$ to $y$,  and  $\delta(x,y) $ is the distance moved by the motor when $X$ jumps from $x$ to $y$.   We follow the usual convention that  $n_{\alpha}>0$ when the system binds particles  and   $n_{\alpha}<0$ when the system releases particles \cite{callen1985thermodynamics}.   The minus sign in front of $f_{\rm mech}$ implies that the mechanical force opposes  the motion of the motor when it moves forward.   
 
\section{Estimators of $\dot{s}$ that rely on the measurement of a single current} \label{sec:3} 
We aim to infer $\dot{s}$ based on the measurements   of  a single stochastic current  of the form  
\begin{equation}
J(t) := \sum_{x,y\in\mathcal{X}}c_{x,y}J_{x\rightarrow y}(t),
\end{equation}
where $c_{x,y}\in\mathbb{R}$; observe that the stochastic entropy production $S$, as defined in (\ref{eq:Stot}), is a specific example of a current.    If $J=cS$, with $c$ a constant, then this is equivalent to an  estimation problem with complete information, while otherwise, this is an estimation problem with partial information.   Without loss of generality, we assume in what follows that $\langle J(t)\rangle/t = \overline{j}>0$.

\subsection{Estimators of dissipation: general definitions}   \label{sec:31}

 Given $n_{\rm s}$ independent  realisations $J^{(i)}(t)$ of  a current $J(t)$, with $i=1,2,\ldots,n_{\rm s}$, an estimator  $\mathcal{F}(\vec{J}^t_0)$ of $\dot{s}$ is a  functional  defined on  the set of trajectories $\vec{J}^t_0 = ([J^{(1)}]^t_0, [J^{(2)}]^t_0,\ldots, [J^{(n_{\rm s})}]^t_0)$.      
 
 The bias of the estimator $\mathcal{F}$ is its average value 
  \begin{equation}
 \langle \mathcal{F}(\vec{J}^t_0) \rangle = \hat{s} +  g(n_{\rm s}), \label{eq:ineq}
 \end{equation}  
 where 
   \begin{equation} 
 \hat{s}= \lim_{n_{\rm s}\rightarrow \infty} \langle \mathcal{F}(\vec{J}^t_0) \rangle
  \end{equation}  
is the bias  of the estimator in the asymptotic limit of an infinite number of realisations, and 
   $g(n_{\rm s})$ is a function that  determines how the bias depends on $n_{\rm s}$; note that $g(n_{\rm s})\rightarrow 0$ for $n_{\rm s}\gg 1$.   An estimator is unbiased in the limit of an infinite number of realisations when  $\hat{s}=\dot{s}$, and it is an unbiased estimator, tout court,  when both  $\hat{s}=\dot{s}$ and  $g=0$.

In what follows, we   define three estimators  of   $\dot{s}$, namely, $\mathcal{F}_{\rm FPR}$ based on the FPR, $\mathcal{F}_{\rm TUR}$ based on the TUR, and $\mathcal{F}_{\rm KL}$ based on the Kullback-Leibler divergence.    As we discuss, all the estimators capture a finite fraction of the dissipation, i.e. 
\begin{equation}
\hat{s} \leq \dot{s},
\end{equation}
but the equality is attained in different circumstances.   Consequently, all the functionals we introduce are useful as  estimators of dissipation, but they capture dissipation in  different scenarios.

\subsection{An estimator based on the first-passage ratio $\hat{s}_{\rm FPR}$  (FPR) }  \label{sec:31} 

\subsubsection{Definition.}
We define  an estimator $\mathcal{F}_{\rm FPR}$ of $\dot{s}$  for which it holds that 
\begin{equation}
\lim_{n_{\rm s}\rightarrow\infty}\mathcal{F}_{\rm FPR} = \hat{s}_{\rm FPR} ,
\end{equation}
where  $\hat{s}_{\rm FPR} $ is the first-passage ratio Eq.~(\ref{eq:FPR}), with 
\begin{equation}
T_J = {\rm inf}\left\{t>0: J(t)\notin (-\ell_-, \ell_+)\right\} \label{eq:FPTime}
\end{equation} 
and 
\begin{equation}
p_- = \mathbb{P}\left(J(T_J)\leq -\ell_-\right). 
\end{equation}

Since an estimator $\mathcal{F}$ is defined on $n_{\rm s}$ trajectories of the current $J$, we consider the first-passage times 
\begin{equation}
T^{(i)}_J = {\rm inf}\left\{t>0: J^{(i)}(t)\notin (-\ell_-,\ell_+)\right\} \quad {\rm with} \quad i\in\left\{1,2,\ldots,n_{\rm s}\right\}, \label{eq:defj}
\end{equation} 
associated with the  different trajectories $[J^{(i)}]^t_{0}$ of the current, and  the decision variables 
\begin{equation}
D^{(i)} = \left\{ \begin{array}{ccc} 1&{\rm if}&J^{(i)}(T^{(i)})\geq \ell_+, \\ -1&{\rm if}&J^{(i)}(T^{(i)})\leq -\ell_-,  \end{array}\right. \label{eq:dj}
\end{equation}
that determine for each trajectory whether the current  crosses  first the positive threshold ($D^{(i)}=1$) or  crosses first  the negative threshold ($D^{(i)}=-1$).

Using $T^{(i)}$ and $D^{(i)}$, we  define, assuming $\langle J(t)\rangle>0$, the estimator
 \begin{equation}
 \mathcal{F}_{\rm FPR} :=  \frac{\ell_+}{\ell_- }\frac{1}{\hat{T}_J} \left|\ln \hat{p}_-\right|, \label{eq:FFPR}
\end{equation}
where 
\begin{equation}
\hat{T}_J = \frac{1}{n_{\rm s}}\sum^{n_{\rm s}}_{i=1}T^{(i)}_J  \label{eq:estimatedTJ}
\end{equation}
is the empirical mean of the first-passage time, and 
\begin{equation}
 \hat{p}_-  =\frac{1}{n_{\rm s}}  {\rm max} \left\{\sum^{n_{\rm s}}_{i=1}\delta_{D^{(i)},-1},1\right\} \label{eq:estimatedpM}
\end{equation}
is the empirical estimate of the splitting probability $p_-$.       Analogously, we can define the estimator $ \mathcal{F}_{\rm FPR}$ for   $\langle J(t)\rangle<0$.

\subsubsection{Bias.} 
For stationary Markov jump processes, it holds that~\cite{roldan2015decision, neri2021universal} 
\begin{equation}
\frac{\hat{s}_{\rm FPR}}{\dot{s}} \leq 1+o_{\ell_{\rm min}}(1),\label{eq:FPRBound}
\end{equation}
where  we added $o_{\ell_{\rm min}}(1)$ to indicate that the inequality holds asymptotically for large thresholds $\ell_{\rm min} = {\rm min}\left\{\ell_+,\ell_-\right\}$; here $o_{\ell_{\rm min}}(1)$ represents the small-o notation, i.e., a function that converges to zero for large values of      $\ell_{\rm min}$.

The equality in Eq.~(\ref{eq:FPRBound}) is attained when $J=cS$, with $c$ a constant~\cite{roldan2015decision, neri2021universal}, while for currents that are not proportional to $S$ the estimate is in general biased.

Now, let us  discuss how the bias depends on the number of realisations in the process.   For small values of $n_{\rm s}$ it holds that 
\begin{equation}
\hat{p}_- \approx \frac{1}{n_{\rm s}},
\end{equation}
as the number of events with $D^{(i)}=-1$  is negligible.      Consequently, for  $n_{\rm s}  \ll 1/p_-$, $\langle  \mathcal{F}_{\rm FPR}  \rangle$  grows logarithmically in $n_{\rm s}$, i.e.,
\begin{equation}
\Big\langle  \mathcal{F}_{\rm FPR} \Big \rangle  \approx  \frac{\ell_+}{\ell_-}\frac{1}{\langle T_J\rangle} \ln n_{\rm s} \leq  \hat{s}_{\rm FPR} .\label{eq:FPRAv}
\end{equation}
Notice that in Eq.~(\ref{eq:FPRAv}), for our convenience, we have assumed $\langle 1/\hat{T}_J\rangle\approx 1/\langle  T_J\rangle$, as this will not affect the scaling arguments in $n_{\rm s}$.   At  larger values $n_{\rm s}\gg 1/p_-$,  the mean value $\langle  \mathcal{F}_{\rm FPR}  \rangle$  saturates to 
\begin{equation}
\langle  \mathcal{F}_{\rm FPR}  \rangle \approx\hat{s}_{\rm FPR}  .\label{eq:FPREstim}
\end{equation} 
Combining Eq.~(\ref{eq:FPRAv}) with Eq.~(\ref{eq:FPR}), and considering that we assumed $\overline{j}>0$, we obtain   
\begin{equation}
\langle   \mathcal{F}_{\rm FPR}\rangle = \left\{\begin{array}{ccc}  \frac{\ell_+}{\ell_-}\frac{1}{\langle T_J\rangle} \ln n_{\rm s}  &{\rm if}&n_{\rm s}\ll 1/p_- , \\  \frac{\ell_+}{\ell_-}\frac{1}{\langle T_J\rangle} |\ln p_- |&{\rm if}&n_{\rm s}\gg 1/p_- .\end{array}\right. \label{eq:29}
\end{equation} 
Equation~(\ref{eq:29}) implies that around $n_{\rm s}\approx 1/p_-$  the bias of the estimator saturates to its asymptotic value, and the bias grows logarithmically  at smaller values  of $n_{\rm s}$.

For the case when  $J=S$,  martingale theory for $e^{-S}$   provides  exact expressions for $p_-$ in terms of the thresholds $\ell_-$ and $\ell_+$ \cite{chetrite2011two, neri2017statistics, neri2019integral}.   As shown in \ref{App:a}, for Markov jump  processes near equilibrium ($\dot{s}\approx 0$),   
\begin{equation}
p_- = \frac{1-e^{-\ell_+}}{e^{\ell_-}-e^{-\ell_+}},
\end{equation} 
so that we require  a number 
\begin{equation}
n_{\rm s}\gg  \frac{e^{\ell_-}-e^{-\ell_+}}{1-e^{-\ell_+}} \sim e^{\ell_-}
\end{equation}
of samples, 
which grows exponentially in the depth of the negative threshold $\ell_-$.     For process that are governed far from thermal equilibrium, we obtain instead (see   \ref{App:a})
\begin{equation}
p_- \leq  \frac{1}{e^{ {\rm max}\left\{\ell_-,\Delta s\right\}}  -  e^{-\ell_+}}  \label{eq:pmm}
\end{equation}  
where  
\begin{equation}
\Delta s = {\rm inf} |S(t)-\lim_{\epsilon\rightarrow 0^+}S(t-\epsilon)|
\end{equation}
is the infimum of the jump sizes in $S(t)$.     Equation~(\ref{eq:pmm})
implies that the number of samples  required grows as
\begin{equation}
n_{\rm s}\gg e^{ {\rm max}\left\{\ell_-,\Delta s\right\}}  -  e^{-\ell_+} \sim e^{ {\rm max}\left\{\ell_-,\Delta s\right\}},
\end{equation}
which is exponential in either the depth of the negative threshold $\ell_-$ or the minimal jump size $\Delta s$, depending on which one is larger.   Hence, for processes that are governed far from thermal equilibrium, the number of samples required is independent of the thresholds, but instead depends on the jumps size $\Delta s$.

\subsection{An estimator based on the thermodynamic uncertainty ratio  $\hat{s}_{\rm TUR}$ (TUR)}  \label{sec:32} 
\subsubsection{Definition.}
We define an estimator $\mathcal{F}_{\rm TUR}$ of $\dot{s}$  for which it holds that 
\begin{equation}
\lim_{n_{\rm s}\rightarrow\infty}\mathcal{F}_{\rm TUR} = \hat{s}_{\rm TUR} ,
\end{equation}
where  $\hat{s}_{\rm TUR} $  is the thermodynamic uncertainty ratio (\ref{eq:TUR}).   The corresponding estimator $\mathcal{F}_{\rm TUR}$, defined on a  finite number $n_{\rm s}$ of realisations of $X$, is 
\begin{equation}
\mathcal{F}_{\rm TUR} := \frac{2}{t} \frac{\left(\sum^{n_{\rm s}}_{i=1}J^{(i)}(t)\right)^2}{n_{\rm s}\sum^{n_{\rm s}}_{i=1}\left(J^{(i)}(t)\right)^2 - \left(\sum^{n_{\rm s}}_{i=1}J^{(i)}(t)\right)^2  }.
\end{equation}   

\subsubsection{Bias.}
  For stationary Markov jump processes, the TUR-relation  implies that the bias \cite{barato2015thermodynamic, gingrich2016dissipation, pietzonka2017finite, horowitz2017proof}
\begin{equation}
\frac{\hat{s}_{\rm TUR}}{\dot{s}}\leq 1,\label{eq:ineqTUR}
\end{equation}  
and hence also the TUR captures a finite fraction of the total dissipation.  

The equality in Eq.~(\ref{eq:ineqTUR}) is attained by processes near equilibrium.  On the other hand, for processes  driven away far from thermal equilibrium, the bias in $\mathcal{F}_{\rm TUR}$ is large, i.e., $\hat{s}_{\rm TUR} \ll \dot{s}$ \cite{pigolotti2017generic, busiello2019hyperaccurate}.       Improvements on the TUR exist, but these  rely on measurements of both currents and time-integrated empirical measures~\cite{shiraishi2021optimal, PhysRevX.11.041061}, and thus require more information about the process $X$.    

To determine the dependence of   $\langle \mathcal{F}_{\rm TUR} \rangle$ on  $n_{\rm s}$, we rely on the central limit theorem, from which it follows that 
\begin{equation}
\langle \mathcal{F}_{\rm TUR} \rangle  = \hat{s}_{\rm TUR} + O\left(\frac{1}{\sqrt{n_{\rm s}}}\right),  \label{eq:finiteTUR}
\end{equation}
where $O$ denotes the big-O notation.   The correction term in Eq.~(\ref{eq:finiteTUR})  is due to sample-to-sample fluctuations in the current.

\subsection{An estimator based on the Kullback-Leibler divergence $\hat{s}_{\rm KL}$}  \label{sec:33}  

\subsubsection{Definition.}We define an estimator  $\mathcal{F}_{\rm KL}$ of $\dot{s}$ for which it holds that 
\begin{equation}
\lim_{n_{\rm s}\rightarrow \infty}\mathcal{F}_{\rm KL} = \hat{s}_{\rm KL},
\end{equation}
where $\hat{s}_{\rm KL}$ is defined in Eq.~(\ref{eq:KL}).

Consider the set $\mathcal{J}$ of all possible jump sizes  $\Delta J(t) = J(t)-\lim_{\epsilon\rightarrow 0^+}J(t-\epsilon)$ in  $J(t)$, excluding the   $\Delta J(t)=0$ case as it does not count as a jump.    Let $N^{(i)}_{j}(t)$ denote the number of   $\Delta J^{(i)}(t) = j$ occurrences  in the trajectory $[J^{(i)}]^t_0$.    Since we assumed that the process is reversible, see Eq.~(\ref{eq:reversible}), it holds that $-j\in\mathcal{J}$ whenever $j\in\mathcal{J}$.     Therefore, we   define the estimator 
\begin{equation}
\mathcal{F}_{\rm KL}(t) :=   \sum_{j\in\mathcal{J}} \frac{\sum^{n_{\rm s}}_{i=1}N^{(i)}_{j}(t)}{t\: n_{\rm s}} \ln \frac{\sum^{n_{\rm s}}_{i=1}N^{(i)}_{j}(t)}{{\rm max}\left\{\sum^{n_{\rm s}}_{i=1}N^{(i)}_{-j}(t),1\right\}} \label{eq:KLx}
\end{equation} 
that estimates the entropy production  rate assuming that the process is Markovian;   $\mathcal{F}_{\rm KL}(t)$ is closely related to the   {\it apparent} entropy production considered in Refs.~\cite{uhl2018fluctuations, ehrich2021tightest}.       In Eq.~(\ref{eq:KLx}), it should be understood that $0\log(0) = 0$.

\subsubsection{Bias.} It holds that \cite{uhl2018fluctuations, ehrich2021tightest}
\begin{equation}
\frac{\hat{s}_{\rm KL}}{\dot{s}} \leq 1,
\end{equation}
which follows from applying Jensen's inequality to the definition of $\dot{s}$, which states that $\dot{s}$ is a Kullback-Leibler divergence.    The equality is attained when  the current $J$ is Markovian and a linear combination of all independent currents in $X$.

We discuss how the bias $\langle \mathcal{F}_{\rm KL}\rangle$ depends on $n_{\rm s}$.    Let us assume that $\langle J\rangle>0$, $n_{\rm s}t$ is small, and $\dot{s}$  is large enough, so that $N^{(i)}_{-j} =0$ for $j>0$.   In this case,    
\begin{equation}
\fl \langle \mathcal{F}_{\rm KL}(t) \rangle \approx \frac{N(t)}{t\:n_{\rm s}}  \sum_{j\in \mathcal{J}\cap \mathbb{R}^+} w_{j} \ln w_{j} +   \frac{N(t)}{t\:n_{\rm s}} \ln N(t),\label{eq:KLx}
\end{equation}  
where 
\begin{equation}
N(t) = \sum_{j\in\mathcal{J}}\sum^{n_{\rm s}}_{i=1}N^{(i)}_{j}(t) 
\end{equation}
is the total number of jumps observed, and 
\begin{equation}
w_{j} = \frac{\sum^{n_{\rm s}}_{i=1}N^{(i)}_{j}(t)}{N(t)}
\end{equation}
is the fraction of jumps of size  $j$.  Setting $N(t)/t \approx   \dot{n} \:  n_{\rm s}$, we obtain that 
\begin{equation}
 \langle \mathcal{F}_{\rm KL}(t) \rangle    \approx \dot{n}\:   \sum_{j\in \mathcal{J}\cap \mathbb{R}^+} w_{j} \ln w_{j} +   \dot{n}  \ln  \left( \dot{n} \:  n_{\rm s} t\right).
\end{equation}
Hence, analogous to the FPR, the bias of  $\langle \mathcal{F}_{\rm KL}(t) \rangle$ grows logarithmically in the number of samples $n_{\rm s}$ until it reaches its asymptotic value $\hat{s}_{\rm KL}$, and therefore $\langle \mathcal{F}_{\rm KL}(t) \rangle$  takes a form similar to Eq.~(\ref{eq:29}).

In what follows, we compare estimation of dissipation with the three estimators $\mathcal{F}_{\rm FPR}$, $\mathcal{F}_{\rm TUR}$ and $\mathcal{F}_{\rm KL}$, in two paradigmatic examples of nonequilibrium processes.   As discussed, this boils down to determining $\hat{s}/\dot{s}$ and $g(n_{\rm s})$, as defined in Eq.~(\ref{eq:ineq}), for these three estimators.

\section{Random walker on a two-dimensional lattice}  \label{sec:4}
We  use first-passage processes for currents to infer the rate  at which a random walker on a two-dimensional lattice produces entropy.     In particular, we compare estimates based on   $ \mathcal{F}_{\rm FPR}$ with those based on $ \mathcal{F}_{\rm TUR}$ and $\mathcal{F}_{\rm KL}$.

In Sec.~\ref{sec:41} we define the model.    Subsequently, we analyse in Sec.~\ref{sec:42} the bias in the estimator $\mathcal{F}_{\rm FPR}$     for currents that are not proportional to $S$, and in  Secs.~\ref{sec:43}  and \ref{sec:44} we analyse the effect of finite thresholds  and a finite number of realisations, respectively, on the bias of $\mathcal{F}_{\rm FPR}$.

\subsection{Model definition}\label{sec:41}
We consider a two-dimensional hopping process $X = (X_1,X_2)$, with   $X_1,X_2\in\left\{1,2,\ldots,n\right\}$  the coordinates of a random walker on a two-dimensional  lattice of length $n$ with periodic boundary conditions [see Fig.~\ref{fig:examples}(a) for an illustration].   The coordinates evolve according to 
\begin{equation}
{\rm d}X_i(t) =  {\rm d}N^{+}_i(t) -  {\rm d}N^{-}_i(t),  \quad i\in \left\{1,2\right\},\label{eq:randomwalkX1}
\end{equation} 
where $N^+_{i}$ and $N^-_i$ are two counting processes with rates $k^+_i$ and $k^-_i$, respectively, and the boundary conditions are implemented by setting $X_i \pm n = X_i$.

We parameterise the rates as 
\begin{equation}
k^+_1 = \frac{e^{\nu /2}}{4\cosh(\nu /2)}, \quad k^-_1 =\frac{e^{-\nu  /2}}{4\cosh(\nu /2)},\label{eq:KPar}
\end{equation}
and 
\begin{equation}
k^+_2 = \frac{e^{\nu \rho /2}}{4\cosh(\nu  \rho/2)}, \quad k^-_2 = \frac{e^{-\nu \rho/2}}{4\cosh(\nu \rho/2)},\label{eq:KPar2}
\end{equation}
so that  the total rate
\begin{equation}
k_{\rm total} = k^+_1 + k^-_1 + k^+_2 + k^-_2 = 1.
\end{equation}    
Notice that the total rate fixes the characteristic time scale of the process.
We  assume, without loss of generality, that $\nu\geq 0$ and $\rho\geq 1$.

The stochastic entropy production, as defined by (\ref{eq:Stot}), is in the present example given by
\begin{equation}
S(t) = \nu  X_1(t) + \nu  \rho  X_2(t),  \label{eq:SEx}
\end{equation}
as $p_{\rm ss}(x) = 1/n^2$ independent of $x$. 
Consequently, the entropy production rate 
\begin{equation}
\dot{s} =  \nu  (k^+_1-k^-_1) +  \nu \rho(k^+_2-k^-_2). \label{eq:sigmaxM}
\end{equation} 
From Eqs.~(\ref{eq:SEx}) and (\ref{eq:sigmaxM}) it follows that $\nu$ is the affinity associated with $X_1$ and $\rho$ is the ratio between the affinities of  $X_2$ and $X_1$.    From a geometric point of view,  the stochastic process $S$ in Eq.~(\ref{eq:SEx})  can be represented as a vector in the  $(X_1,X_2)$-plane.    The parameter $\rho$  fixes the direction of this vector and $\nu$ its magnitude.

Without loss of generality, we can  parametrise  stochastic currents  in this model by 
\begin{equation} 
J(t) = c(1-\Delta ) X_1(t) + c(1+\Delta) X_2(t).   \label{eq:JM}
\end{equation}       
Since the FPR and TUR are independent of $c$, we  set  $c=1$.  If 
\begin{equation}
\Delta =  \frac{\rho-1}{\rho+1}, \label{eq:DeltaAst}
\end{equation}
then  $J(t)$ is proportional to $S(t)$, which is independent of $\nu$.     

In the present model, the currents $J$    are Markovian.    This is a peculiar feature of the random walk model as, in general, currents in Markov jump processes are nonMarkovian.

\begin{figure}[t!]
\centering
\includegraphics[width=0.65\textwidth]{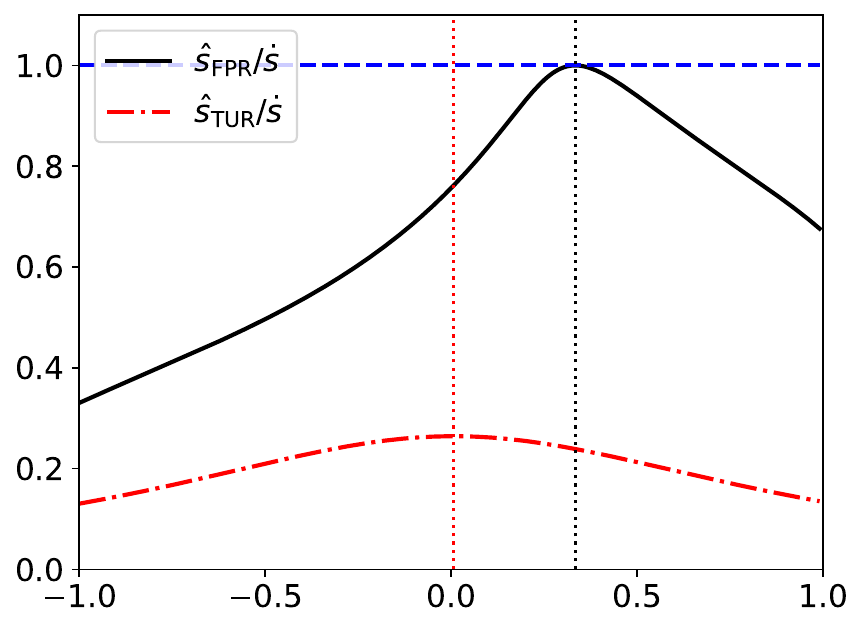}
  \put(-147,-10){\large$\Delta$} 
\caption{ {\it Comparing the  bias of FPRs and TURs  evaluated on currents in a random walk process on a two-dimensional lattice}.    The ratios  $\hat{s}_{\rm FPR}/\dot{s}$  [in the limit of   $\ell_{\rm min}\rightarrow \infty$, given by Eq.~(\ref{eq:hatsmfrp})]  and  $\hat{s}_{\rm TUR}/\dot{s}$ [Eq.~(\ref{eq:Stur})] are plotted as a function of the parameter  $\Delta$ that determines the  current $J$.       The vertical dotted lines determine the values of $\Delta$  at which  $\hat{s}_{\rm FPR}$ and $\hat{s}_{\rm TUR}$ reach their maximum, which are given by $1/3$ [according to Eq.~(\ref{eq:DeltaAst})] and  $0.0067$  [Eq.~(\ref{eq:DeltaTUR})], respectively.     The rates $k^+_1$, $k^-_1$, $k^+_2$, and $k^-_2$ are parametrised according to Eqs.~(\ref{eq:KPar}) and (\ref{eq:KPar2}), with $\rho=2$ and $\nu=5$; the corresponding value of $\dot{s}\approx 7.5 k_{\rm total}$.    The ratio $\hat{s}_{\rm TUR}$ is evaluated at $t=20/k_{\rm total}$.     The dashed, blue, horizontal  line  is a guide to the eye that  indicates the value $1$ corresponding to optimal thermodynamic inference.} \label{fig:sEstim}
\end{figure}  
 
\begin{figure}[t!]
\centering
\includegraphics[width=0.65\textwidth]{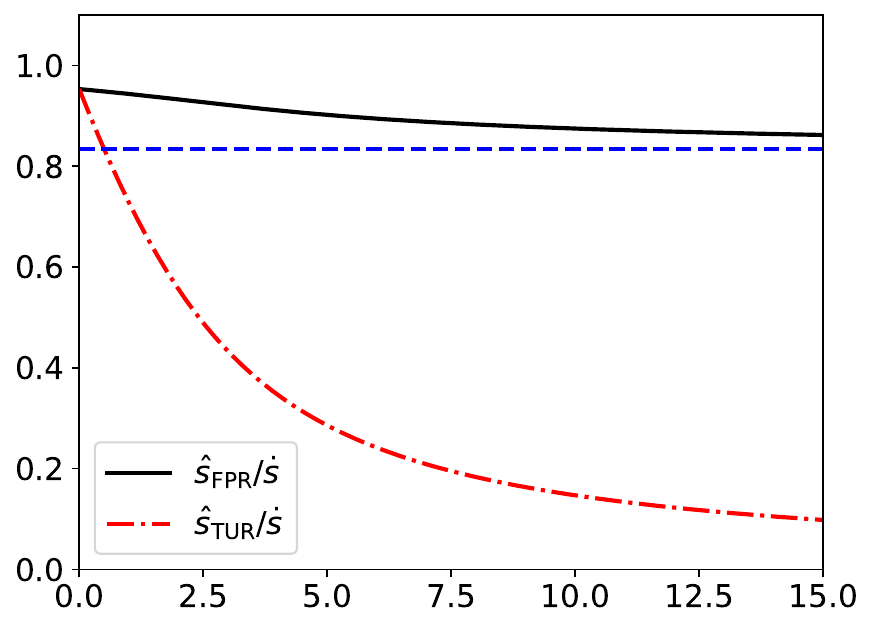}
  \put(-165,-10){\large$\dot{s}/k_{\rm total}$} 
\caption{ {\it  The bias in the FPR and TUR  as a function of the rate of dissipation.}      The ratios  $\hat{s}_{\rm FPR}/\dot{s}$ (in the limit of   $\ell_{\rm min}\rightarrow \infty$) and  $\hat{s}_{\rm TUR}/\dot{s}$ are plotted as a function of the rate of dissipation   $\dot{s}/k_{\rm total}$  for the current with $\Delta=0.6$ in the random walk model defined on a two-dimensional lattice.    The rates $k^+_1$, $k^-_1$, $k^+_2$, and $k^-_2$ are parametrised according to Eqs.~(\ref{eq:KPar}) and (\ref{eq:KPar2}), with $\rho=2$, while the parameter  $\nu$ depends on the rate $\dot{s}$ of dissipation.    The horizontal dashed line denotes the asymptotic limit  $\lim_{\nu\rightarrow \infty}\hat{s}_{\rm FPR}$ given by Eq.~(\ref{eq:sfprs}).  } \label{fig:sEstim2}
\end{figure}

\subsection{Entropy production estimation with currents that are not proportional to $S$}\label{sec:42}
When $J$ is proportional to $S$, then   $\hat{s}_{\rm FPR}$ is unbiased in the limit of large thresholds, and hence in this case the FPR is an optimal estimator of dissipation.   
However, for currents that are not proportional to $S$,  $\hat{s}_{\rm FPR}$ is biased.    We analyse here the effect of the parameter $\Delta$ that determines the current $J$ on the biases in $\hat{s}_{\rm FPR}$ (in the limit of large thresholds), $\hat{s}_{\rm TUR}$ and $\hat{s}_{\rm KL}$.

   In the limit of large thresholds, the mean first-passage time (see Eq.~(17) of Ref.~\cite{neri2021universal})
\begin{equation}
\langle T_J\rangle  = \frac{\ell_+}{\overline{j}} = \frac{\ell_+}{(1-\Delta)(k^+_1 -k^-_1)+ (1+\Delta)(k^+_2 -k^-_2 )}
\end{equation} 
and the splitting probability (see~Proposition 7 of Appendix E of \cite{neri2021universal})
  \begin{eqnarray}
 p_- =e^{\ell_- z^\ast} \label{eq:P-x},  
\end{eqnarray}   
  where    $z^\ast<0$ is the nonzero solution to 
  \begin{equation}
\fl (1-e^{(1-\Delta)z^\ast})k^+_1 + (1-e^{-(1-\Delta)z^\ast})k^-_1
+   (1-e^{(1+\Delta)z^\ast})k^+_2 + (1-e^{-(1+\Delta)z^\ast})k^-_2 = 0. \label{eq:zAstSolv}
\end{equation}   
Hence, for  $\ell_{\rm min} = {\rm min}\left\{\ell_-,\ell_+\right\}\gg 1$ the ratio Eq.~(\ref{eq:FPR}) reads
\begin{equation}
\frac{\hat{s}_{\rm FPR}}{\dot{s}} = \frac{|z^\ast|}{\nu}    \frac{ (1-\Delta)(k^+_1 -k^-_1)+ (1+\Delta)(k^+_2 -k^-_2 )}{ k^+_1-k^-_1 +   \rho(k^+_2-k^-_2)}, \label{eq:hatsmfrp}
\end{equation}
which is independent of the ratio $\ell_+/\ell_-$.

Figure~\ref{fig:sEstim}  plots $\hat{s}_{\rm FPR}$ as a function of $\Delta$ and shows graphically   that  $\hat{s}_{\rm FPR} \leq \dot{s}$, in correspondence with the general bound (\ref{eq:FPRBound}).     Moreover, the equality  $\hat{s}_{\rm FPR}=\dot{s}$ is attained when $\Delta$ is  given by Eq.~(\ref{eq:DeltaAst}),  as indicated by the vertical, black, dotted line in Fig.~\ref{fig:sEstim}, and hence $\hat{s}_{\rm FPR}$ is an unbiased estimator when  $J(t)=S(t)$.    Indeed,  this can be verified analytically, as 
\begin{equation}
z^\ast = -\frac{\nu(\rho+1)}{2}\label{eq:zast}
\end{equation}
 solves the Eq.~(\ref{eq:zAstSolv}) when $\Delta$ is given by Eq.~(\ref{eq:DeltaAst}).  Substituting the values of $\Delta$ and $z^\ast$ given by (\ref{eq:DeltaAst}) and (\ref{eq:zast}) in Eq.~(\ref{eq:hatsmfrp}), we recover (\ref{eq:sigmaxM}), and hence we obtain in this case that   $\hat{s}_{\rm FPR}=\dot{s}$.

Let us now compare the biases of $\hat{s}_{\rm FPR}$ and $\hat{s}_{\rm TUR}$.    A direct computation of $\overline{j}$ and $\sigma^2_{J(t)}$ for a random walker on a two-dimensional lattice gives  
\begin{equation}
\fl \frac{\hat{s}_{\rm TUR}}{\dot{s}} = \frac{1}{\nu}  \frac{ 2\left[ (1-\Delta)(k^+_1 -k^-_1)+ (1+\Delta)(k^+_2 -k^-_2 )\right]^2}{[  k^+_1-k^-_1 +   \rho(k^+_2-k^-_2)] [(1-\Delta)^2(k^+_1 +  k^-_1) + (1+\Delta)^2 (k^+_2 +  k^-_2)]}.  \label{eq:Stur}  
\end{equation}  
Figure~\ref{fig:sEstim} plots $\hat{s}_{\rm TUR}/\dot{s}$ as a function of $\Delta$, demonstrating graphically that  $\hat{s}_{\rm TUR} < \dot{s}$ in correspondence with the bound Eq.~(\ref{eq:ineqTUR}).       In contrast to $\hat{s}_{\rm FPR}$, however, which is unbiased at $\Delta=1/3$, theTUR captures at most about 26\% of the total dissipation for these parameters, even for the optimal choice of current. This is because the TUR systematically underestimates dissipation far from thermal equilibrium, and the parameters used here, for which $\dot{s} \approx 7.5 k_{\rm total}$, correspond to a strongly driven regime.

To demonstrate that at high values of $\dot{s}$   the FPR  estimates  better the entropy production rate than the TUR, we   consider the asymptotic limit of $\nu\gg 1$, corresponding with large dissipation rates $\dot{s}$.      In this limit,    the nonzero solution to Eq.~(\ref{eq:zAstSolv}) is 
\begin{equation}
z^\ast =  \left\{\begin{array}{ccc}  \frac{\nu}{\Delta-1} &{\rm if}& \Delta < \frac{\rho-1}{\rho+1} ,  \\ -\frac{\nu \rho}{1+\Delta} &{\rm  if}&\Delta > \frac{\rho-1}{\rho+1},\end{array}\right.
\end{equation} 
so that   the FPR captures a  nonzero fraction of the dissipation in the limit of large entropy production rates, viz., 
\begin{equation}
\lim_{\nu\rightarrow \infty}\frac{\hat{s}_{\rm FPR}}{\dot{s}} = \left\{ \begin{array}{ccc}  \frac{2}{(1+\rho)(1-\Delta)} &{\rm if}& \Delta < \frac{\rho-1}{\rho+1} , \\  \frac{2 \rho}{(1+\rho)(1+\Delta)}&{\rm  if}&\Delta > \frac{\rho-1}{\rho+1}. \end{array} \right. \label{eq:sfprs}
\end{equation} 
On the other hand,  in the limit of large $\nu$, Eq.~(\ref{eq:Stur}) implies  that  
\begin{equation}
\lim_{\nu\rightarrow \infty} \frac{\hat{s}_{\rm TUR}}{\dot{s}} =  \frac{1}{\nu}\frac{4}{(1+\Delta^2)(1+\rho)} + O(\nu^{-2}) ,
\end{equation}
and hence  in the limit of $\dot{s}\gg 1$ the TUR captures a zero fraction of the total dissipation.      

The fact that $\hat{s}_{\rm FPR}$ captures a nonzero fraction of  the entropy production rate far from thermal equilibrium, while $\hat{s}_{\rm TUR}$ captures in this limit a zero fraction of the entropy production rate, is demonstrated graphically in Fig.~\ref{fig:sEstim2}.

   Another aspect of $\hat{s}_{\rm FPR}$ that is  well illustrated with Fig.~\ref{fig:sEstim} is that the maximum value of $\hat{s}_{\rm FPR}$ is attained when $J=cS$.  Instead,   for $\hat{s}_{\rm TUR}$ the most precise current is determined by 
\begin{equation}
\Delta  =   \frac{\tanh(\nu \rho/2)-\tanh(\nu/2)}{\tanh(\nu \rho/2)+\tanh(\nu/2)}.      \label{eq:DeltaTUR}
\end{equation} 
Hence, the most precise current that maximises $\hat{s}_{\rm TUR}$ is  in general not proportional to the  stochastic entropy production \cite{busiello2019hyperaccurate, busiello2022hyperaccurate}.  
For example, in Fig.~\ref{fig:sEstim}, the most precise current has a value $\Delta\approx 0.0067$, which is different from $\Delta=1/3$ corresponding with $J=S$ (both are indicated by the vertical dotted lines).

Lastly, let us compare $\hat{s}_{\rm FPR}$ with $\hat{s}_{\rm KL}$.  Since $J$ is Markovian, it holds that $\hat{s}_{\rm KL} = \dot{s}$, and hence in the present example estimates based on the Kullback-Leibler divergence are unbiased.      This is a specific feature of the random walk model that does not extend to other models, as  currents in Markov jump processes are in general nonMarkovian, so that,  $\hat{s}_{\rm KL} < \dot{s}$ (see Sec.~\ref{sec:5} for a general example).  For this reason, the random walk  model studied in this section is mainly useful  to compare  entropy production estimates based on  FPRs with those based on TURs.  

\subsection{Effect of finite thresholds}  \label{sec:43}
So far, we have considered  $\hat{s}_{\rm FPR}(\ell_-,\ell_+)$ in the limit of large thresholds  $\ell_+,\ell_-\gg 1$.     We determine now    the effect of finite thresholds on  $\hat{s}_{\rm FPR}(\ell_-,\ell_+)/\dot{s}$.      For finite thresholds,   we  solve numerically a recursion formula to determine $\hat{s}_{\rm FPR}(\ell_-,\ell_+)$, as explained in \ref{App:pM}.  Although there is no conceptual difficulty in considering asymmetric thresholds,  to reduce the number of parameters, we consider symmetric thresholds  $\ell_-=\ell_+=\ell$.

\begin{figure}[h!]
\centering
  \includegraphics[width=0.65\textwidth]{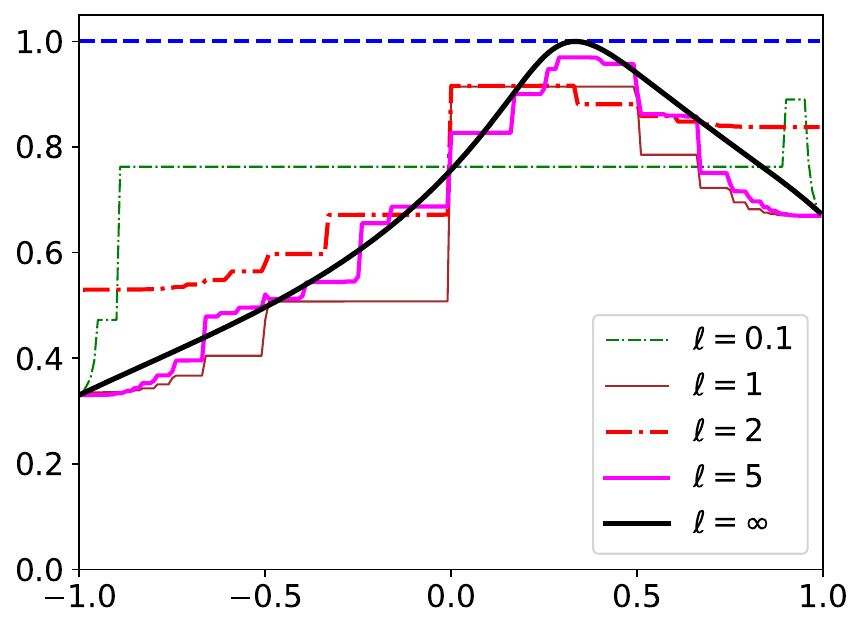}
  \put(-146,-10){\Large$\Delta$} 
    \put(-320,100){\Large$\frac{\hat{s}_{\rm FPR}}{\dot{s}}$} 
\caption{ {\it Effect of finite thresholds on  the estimator $\hat{s}_{\rm FPR}$ (part I). } The function   $\hat{s}_{\rm FPR}/\dot{s}$ is plotted as a function of $\Delta$ for finite values of the thresholds $\ell_-=\ell_+=\ell$.   Parameters are the same as in Fig.~\ref{fig:sEstim}, except that the thresholds are finite, as given in the legend.   The dashed, blue, horizontal  line  is a guide to the eye that  indicates the value $1$ corresponding to optimal inference.} \label{fig:sEstimx}
\end{figure}

Figure~\ref{fig:sEstimx} plots the ratio $\hat{s}_{\rm FPR}(\ell_-,\ell_+)/\dot{s}$  for parameters that are the same as in Fig.~\ref{fig:sEstim}, but now for finite values of $\ell$.         It is evident from  Fig.~\ref{fig:sEstimx}  that even at relatively small values of the thresholds, $\hat{s}_{\rm FPR}(\ell_-,\ell_+)$  estimates dissipation more  accurately than    $\hat{s}_{\rm TUR}$.     For example, for $\ell=0.1$  the estimator $\hat{s}_{\rm FPR}(\ell_-,\ell_+)$  captures for $\Delta\approx 0$ about  75\% of $\dot{s}$, which should be compared with the 25\% captured by the TUR.    

\begin{figure}[h!]
\centering
\subfigure{
\includegraphics[width=0.55\textwidth]{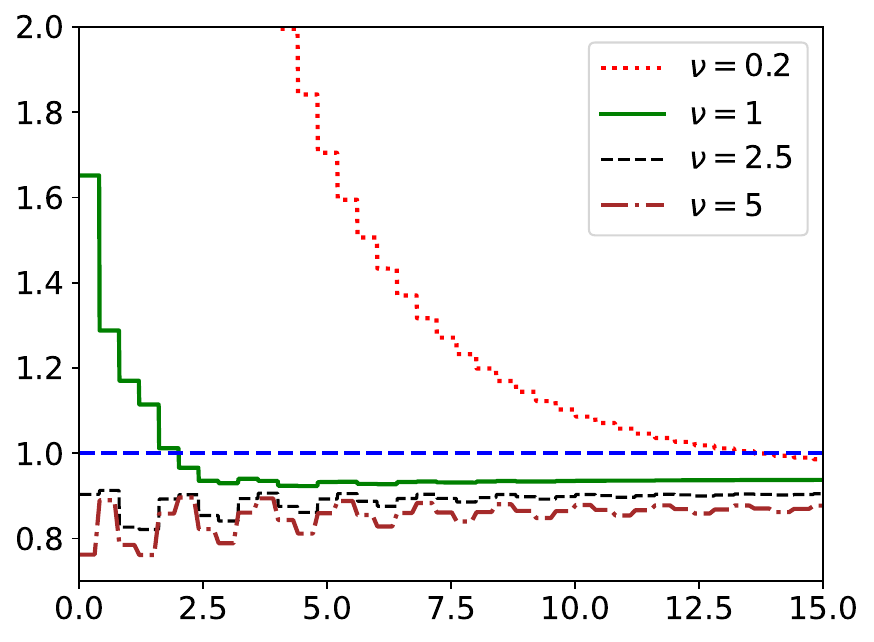}}
  \put(-124,-10){\large$\ell_-$} 
      \put(-275,90){\Large$\frac{\hat{s}_{\rm FPR}}{\dot{s}}$} 
      \hspace{2cm}
 \subfigure{ \includegraphics[width=0.55\textwidth]{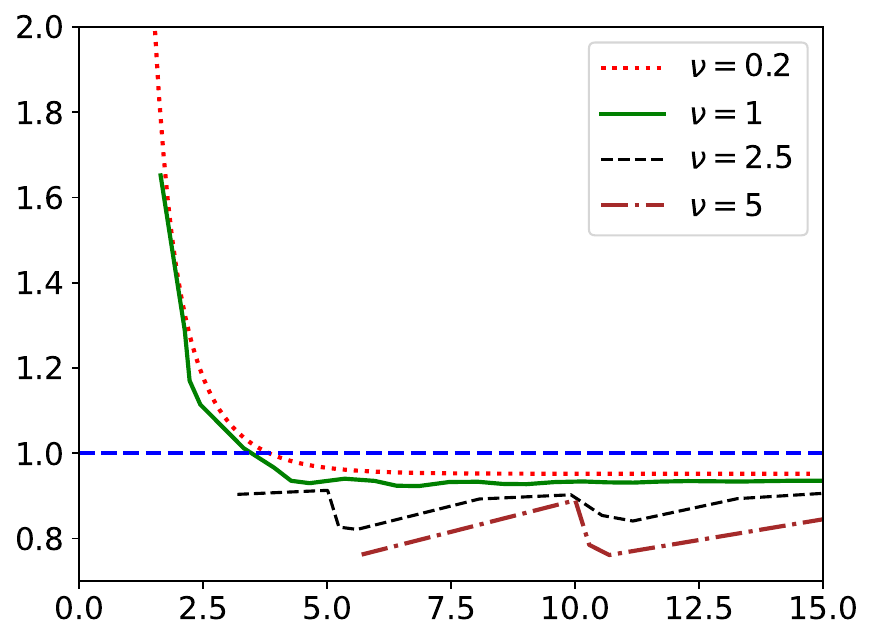}}
    \put(-136,-10){\large$|\ln p_-|$} 
          \put(-275,90){\Large$\frac{\hat{s}_{\rm FPR}}{\dot{s}}$} 
  \caption{ {\it Effect of finite thresholds on  the estimator $\hat{s}_{\rm FP}$ (part II). }     The ratio   $\hat{s}_{\rm FPR}/\dot{s}$ is plotted as a function of $\ell_-$ (top) and $|\ln p_-|$ (bottom) for the current $J$ with $\Delta=0.6$ (defined in Eq.~(\ref{eq:JM})) in the random walk process on a two-dimensional lattice. The rates $k^+_1$, $k^-_1$, $k^+_2$, and $k^-_2$ are parametrised according to Eqs.~(\ref{eq:KPar}) and (\ref{eq:KPar2}), with $\rho=2$ and $\nu=0.2, 1, 2.5, 5$, as given in the legend; the corresponding values of $\dot{s}/k_{\rm total}$ are  $0.05$, $1$, $3.5$, and $7.5$, respectively.    The thresholds are symmetric, $\ell_-=\ell_+$.  The blue, horizontal, dashed line  is a guide to the eye that  indicates the value $1$.   } \label{fig:sEstimxx}
\end{figure}  

In Fig.~\ref{fig:sEstimxx},  we analyse  $\hat{s}_{\rm FPR}(\ell_-,\ell_+)$ as a function of  $\ell$   for a fixed value of $\Delta=0.6$; note that this is a generic example for which $J$ is not proportional to $S$, and hence $\hat{s}_{\rm FPR}$ does not capture all the dissipation in the process, not even in the limit of large thresholds.        The top panel of Fig.~\ref{fig:sEstimxx}    plots      $\hat{s}_{\rm FPR}(\ell_-,\ell_+)/\dot{s}$ as a function of $\ell$ for different values of $\nu$, which is the parameter that determines the hopping rates   according to Eqs.~(\ref{eq:KPar}) and (\ref{eq:KPar2}).      At large values  $\nu\gg 1$ the rate of dissipation is large, and  at small values  $\nu\approx0$  the system is almost at equilibrium.   In Fig.~\ref{fig:sEstimxx}, we  observe that   asymptotically for large $\ell$,  $\hat{s}_{\rm FPR}(\ell_-,\ell_+)/\dot{s}<1$, in accordance with the universal  first-passage bound (\ref{eq:FPRBound}).      However, how fast $\hat{s}_{\rm FPR}(\ell_-,\ell_+)/\dot{s}$ converges to its asymptotic limit depends on $\nu$:  near equilibrium, e.g., for $\nu=0.2$,  large values of $\ell\approx 15$ are required to estimate $\dot{s}$ accurately, while far from thermal equilibrium, e.g., for $\nu>2.5$,  the ratio $\hat{s}_{\rm FPR}$ captures about 90\% of the dissipation for any value of  $\ell$.    Hence, for systems far from thermal equilibrium, $\hat{s}_{\rm FPR}$ estimates dissipation well, even at very small values of the thresholds.  

In order to better understand what value of $\ell$ is required to estimate dissipation, we plot in the bottom panel of Fig.~\ref{fig:sEstimxx}  the ratio  $\hat{s}_{\rm FPR}(\ell_-,\ell_+)/\dot{s}$ as a function of $|\ln p_-|$.       This figure shows that $\hat{s}_{\rm FPR}(\ell_-,\ell_+)$ estimates well the entropy production rate whenever $|\ln p_-|>3$, and hence the relevant asymptotic parameter is $p_-$, rather than the threshold value $\ell$.   Accordingly, for systems near equilibrium, a large threshold is needed to bring $p_-<0.05 = e^{-3}$, while for systems far from thermal equilibrium $p_-<0.05=e^{-3}$ is reached for any value of the thresholds, as described by Eq.~(\ref{eq:pmm}).

\subsection{Effect of a finite number of realisations}  \label{sec:44}
So far, we have analysed $\hat{s}_{\rm FPR}(\ell_-,\ell_+)$, which is the value of the estimator $\mathcal{F}_{\rm FPR}$ when an infinite number of realisations are available.     However, in  experimental situations, we have to deal with a finite amount of data, and hence it is crucial to understand how fast  $\mathcal{F}_{\rm FPR}$ converges to  $\hat{s}_{\rm FPR}(\ell_-,\ell_+)$ as a function of $n_{\rm s}$, the number of realisations of the process $X$.      We address this question here by simulating the process $X$ with a continuous-time Monte Carlo algorithm.

\begin{figure*}[h!]
\centering
\subfigure[$\nu=0.2$]{
\includegraphics[width=0.55\textwidth]{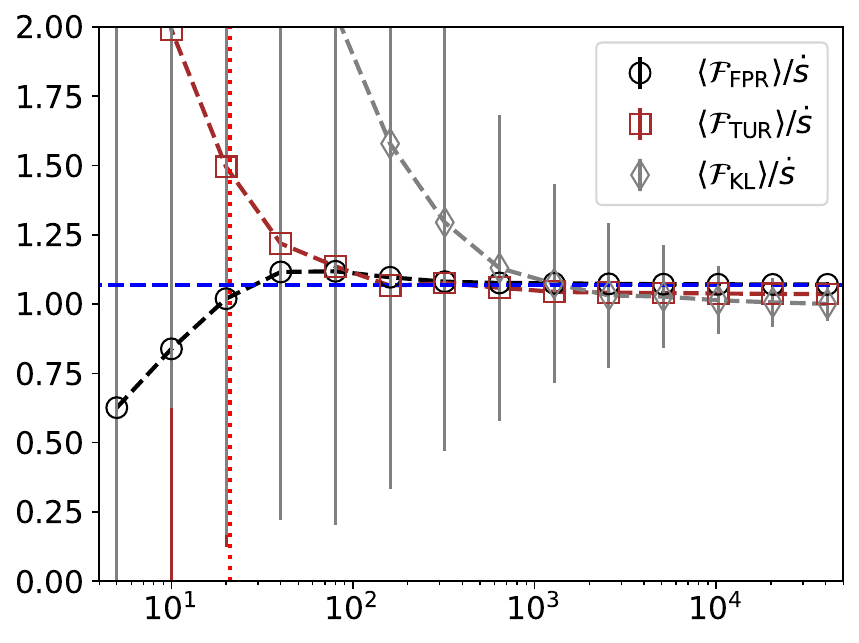}}
  \put(-130,0){\Large$n_{\rm s}$}
\subfigure[$\nu=1$]{
\includegraphics[width=0.55\textwidth]{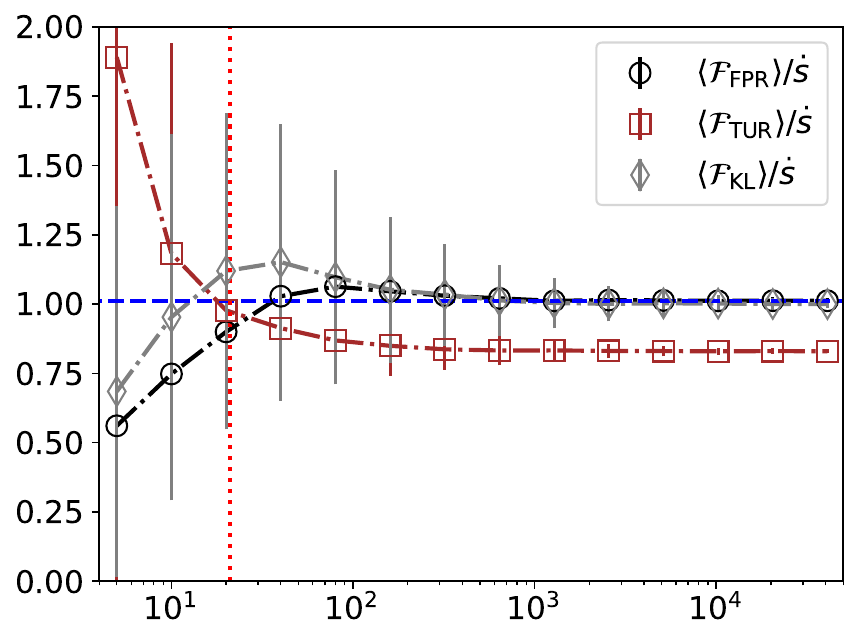}}
  \put(-130,0){\Large$n_{\rm s}$}
   \\
   \subfigure[$\nu=2.5$]{
\includegraphics[width=0.55\textwidth]{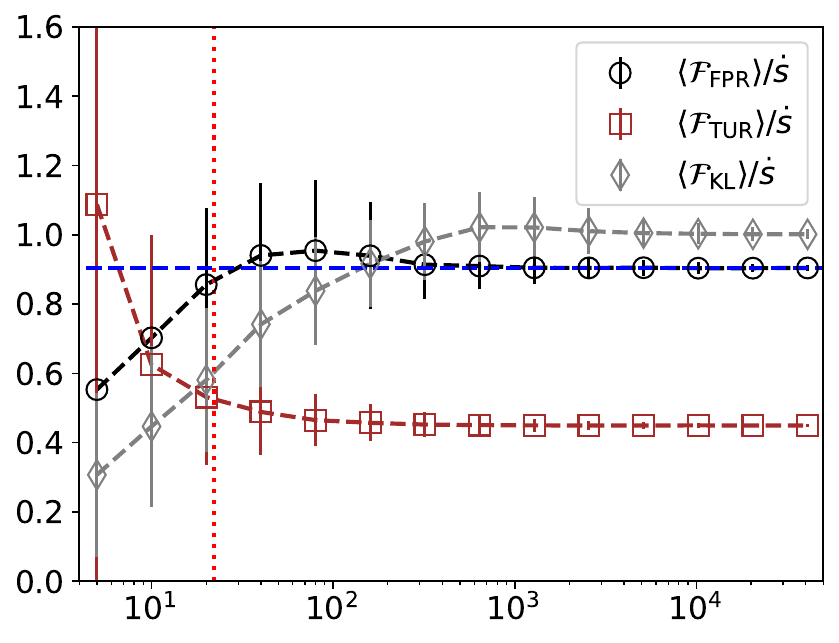}}
  \put(-130,0){\Large$n_{\rm s}$}
    \subfigure[$\nu=5$]{
\includegraphics[width=0.55\textwidth]{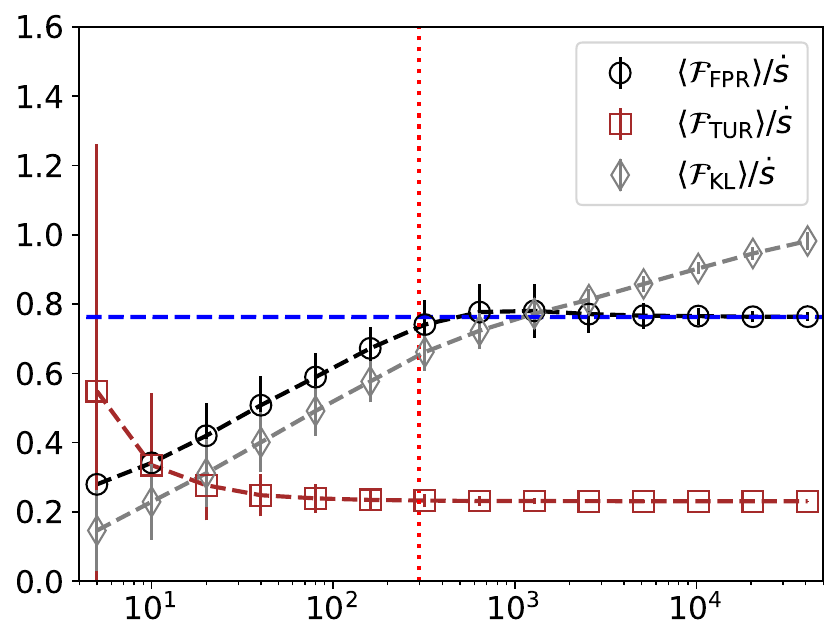}}
  \put(-130,0){\Large$n_{\rm s}$}
  \caption{ {\it Effect of the sample size $n_{\rm s}$ on the bias of estimators of dissipation.}    Circles, squares, and diamonds denote, respectively,   the expected values of $\langle \mathcal{F}_{\rm FPR}\rangle$,  $\langle \mathcal{F}_{\rm TUR}\rangle$ and $\langle \mathcal{F}_{\rm KL}\rangle$ ; error bars denote the standard deviations of $\mathcal{F}_{\rm FPR}$ and $\mathcal{F}_{\rm TUR}$.     Results shown are   for the current $J$ with $\Delta=0.6$ in the random walk process on a two-dimensional lattice.  Different figures represent different values of $\nu$, corresponding with increasingly large rates of  dissipation, ranging from $\dot{s} = 0.05 k_{\rm total}$ when $\nu=0.2$ to $\dot{s} = 7.5 k_{\rm total}$ when $\nu=5$.     The same parameters have been selected as in Fig.~\ref{fig:sEstimxx}, albeit for a fixed  threshold   $\ell_-=\ell_+=\ell$.     The thresholds  are set at $\ell=10.41,1.61,0.01,0.01$ for $\nu=0.2, 1, 2.5, 5$, respectively, and the TUR and the Kullback-Leibler divergence are evaluated at $t=1/k_{\rm total}$.         Markers are empirical averages over $1{\rm e}+3$ runs, each consisting of $n_{\rm s}$ simulated realisations of $X$.   The horizontal, blue, dashed lines denote $\hat{s}_{\rm FPR}/\dot{s}$, and  the vertical, red, dotted  lines denote the value  $n_{\rm s} = 1/p_-$.       } \label{fig:samples}
\end{figure*}  

Figure~\ref{fig:samples} shows $ \mathcal{F}_{\rm FPR}/\dot{s}$ as a function of $n_{\rm s}$ for the same parameters as in Fig.~\ref{fig:sEstimxx} and for a fixed value of $\ell$.   The value of $\ell$ is set to  the smallest threshold value so that  $p_- < 0.05$, in particular, $\ell=10.41,1.61,0.01,0.01$ for $\nu=0.2, 1, 2.5, 5$, respectively; note that for $\nu=2.5$ and $\nu=5$,  we have set $\ell=0.01$, as setting $\ell=0$ would terminate the process immediately.      

We observe two regimes in Fig.~\ref{fig:samples}: (a) when $n_{\rm s}<1/p_-$, then $\langle \mathcal{F}_{\rm FPR}\rangle$  grows logarithmically in $n_{\rm s}$; (b) when $n_{\rm s}>1/p_-$, then $\langle \mathcal{F}_{\rm FPR}\rangle$ saturates to its asymptotic value $\hat{s}_{\rm FPR}$.   To make it easier to differentiate these two regimes, Fig.~\ref{fig:samples} shows the value $n_{\rm s} = 1/p_-$ with the vertical dotted lines.  Hence, we conclude that, in correspondence with Eq.~(\ref{eq:29}), a number $n_{\rm s}\sim 1/p_-$ of samples is required to optimally estimate the rate of dissipation.  The saturation  of $\langle \mathcal{F}_{\rm FPR}\rangle$ at $n_{\rm s}\sim 1/p_-$ has the following intuitive explanation:  to determine the magnitude of  $p_-$, it is necessary to observe the frequency of events that hit the negative threshold first.   In the regime $n_{\rm s}\ll 1/p_-$  the experimenter does not observe events for which the current  hits  the negative threshold first, while for $n_{\rm s}\gg 1/p_-$  the experimenter  observes several such events.     

Comparing estimates of dissipation based on the FPR  with those based on the TUR, we observe that for small values of $\nu$ (or  $\dot{s}$) both estimators are equivalent: their asymptotic  values  for $n_{\rm s}\gg 1$ are almost identical, in correspondence with Fig.~\ref{fig:sEstim2}, and the estimators  converge equally fast to the asymptotic limit, viz., about $n_{\rm s}\sim 10^2$ realisations are required.    On the other hand, for large values of $\nu$ (or $\dot{s}$), the FPR captures asymptotically a finite fraction of the entropy production rate, while the TUR captures  a negligible fraction of the entropy production rate.  However, the small bias of the FPR comes at the expense of a large sample size $n_{\rm s}$  that is required to reach the asymptotic limit (about $n_{\rm s}\sim 10^3$ realisations are required for the FPR compared to $n_{\rm s}\sim 30$ for the TUR).      

Comparing estimations of dissipation based on     $\mathcal{F}_{\rm FPR}$  with those based on $\mathcal{F}_{\rm KL}$, we observe  that the bias in  $\mathcal{F}_{\rm KL}$ is smaller than the bias in $\mathcal{F}_{\rm FPR}$.   This is because in the present model the current is Markovian, and consequently the estimator $\mathcal{F}_{\rm KL}$ is unbiased.     However, the dependence on $n_{\rm s}$ is similar for both estimators, viz., for large values of $\nu$ the  bias grows logarithmically until it reaches its asymptotic value.    

Taken together,  Fig.~\ref{fig:samples} confirms the  Eq.~(\ref{eq:29}), and hence although  far from thermal equilibrium the  FPR   estimates   a finite fraction of the total dissipation, this comes at the expense of a possibly large number of samples $n_{\rm s}\gtrsim 1/p_-$, depending on how small the  splitting probability $p_-$  is.

\section{Inference of the efficiency of molecular motors}  \label{sec:5}
As an application, we  use first-passage processes to estimate the efficiency at which a molecular motor converts chemical fuel into mechanical work.      We assume that the position of the molecular motor, which we denote by $Y$, is observed.  Instead, the internal degrees of freedom, such as the chemical states of the motor heads,  are not known, and hence this is a problem with partial information.       Moreover, in this example the observed current $Y$
 is nonMarkovian, and hence with this model we can quantify the influence of nonMarkovian statistics on entropy production estimates.   In particular, it will be interesting to compare the biases of  $\mathcal{F}_{\rm FPR}$  and $\mathcal{F}_{\rm KL}$.

\subsection{General setup: dissipation and  efficiency of molecular motors}
Consider  a molecular motor that moves on a one-dimensional biofilament at mean velocity $v$ in steps of length $\delta$.    We denote by  $f_{\rm mech}$  the mechanical load felt by the molecular motor  ($f_{\rm mech}>0$ denotes a hindering load, while $f_{\rm mech}<0$ is an assisting load).  When $f_{\rm mech}v>0$,  then the molecular motor consumes chemical fuel in the hydrolysis of adenosine triphosphate (ATP)  into adenosine diphosphate  (ADP) and an inorganic phosphate (P) to perform work on an external cargo.   Following \cite{pietzonka2016universal}, we   define the efficiency of the molecular motor as the ratio
\begin{equation}
\eta := \frac{f_{\rm mech} v}{\dot{w}_{\rm chem}},  \label{eq:wMotor}
\end{equation} 
between the output power $f_{\rm mech}v$ and  the rate $\dot{w}_{\rm chem}$ of consumption of free energy input due to the hydrolysis of ATP into ADP and P.      

The rate of dissipation is the sum of  the rate of entropy change in the chemostats, given by $\dot{w}_{\rm chem}/\mathsf{T}_{\rm env}$, and  in the thermal reservoir, given by $-f_{\rm mech}v/\mathsf{T}_{\rm env}$, viz.,  
\begin{equation}
\dot{s} = \frac{\dot{w}_{\rm chem}}{\mathsf{T}_{\rm env}} -  \frac{f_{\rm mech} v}{\mathsf{T}_{\rm env}}. \label{eq:dotsMotor}
\end{equation}

Combining the Eqs.~(\ref{eq:wMotor}) and (\ref{eq:dotsMotor}), we obtain the relation
\begin{equation}
\dot{s} = \frac{f_{\rm mech} v}{\mathsf{T}_{\rm env}} \left(\frac{1-\eta}{\eta} \right) \label{eq:sEff}
\end{equation}
between the rate of dissipation $\dot{s}$ and the motor efficiency.

  \subsection{A back-of-the-envelope estimate of the efficiency of molecular motors with first-passage processes}\label{sec:52x}
Combining the FPR-inequality Eq.~(\ref{eq:FPRBound})  with  the expression (\ref{eq:sEff}), we obtain the upper bound
\begin{equation}
\eta \leq    \eta_{\rm FPR} = \left(1+   \frac{\mathsf{T}_{\rm env}}{f_{\rm mech} v} \frac{\ell_+}{\ell_-}\frac{ \left|\ln  p_- \right|}{\langle T_Y\rangle} (1+o_{\ell_{\rm min}}(1))\right)^{-1}, \label{eq:bound}
\end{equation}   
on the efficiency of a molecular motor.      The bound Eq.~(\ref{eq:bound}) is the FPR equivalent of the   bound  \cite{pietzonka2016universal},
\begin{equation}
\eta \leq  \eta_{\rm TUR} = \left(1+\frac{\mathsf{T}_{\rm env}}{f_{\rm mech} \delta}\frac{2\langle Y(t)\rangle}{\sigma^2_{Y(t)}}\right)^{-1}
\end{equation} 
that follows from the TUR inequality, given by Eq.~(\ref{eq:ineqTUR}).   Analogously, there is the bound 
\begin{equation}
\eta \leq  \eta_{\rm KL} = \left(1+\frac{\mathsf{T}_{\rm env}}{f_{\rm mech} v} \left( \overline{j}_{\rm f}-\overline{j}_{\rm b}\right) \log \frac{\overline{j}_{\rm f}}{\overline{j}_{\rm b}}\right)^{-1}\label{eq:KLEeta}
\end{equation} 
based on the Kullback-Leibler divergence.

We  estimate $\eta_{\rm FPR}$ based on the experimental data for Kinesin-1 published in Ref.~\cite{kojima1997mechanics}.    For forces $f_{\rm mech}=3 \: {\rm pN}$ and $ [{\rm ATP}] = 10\: \mu {\rm M}$, a number of 2-8\% of backward steps were observed \cite{kojima1997mechanics}.         Hence, if we set the thresholds   
 $\ell_-=\ell_+ = \delta  = 8 \: {\rm nm}$, then $p_-\sim 0.02$, and  we are in the asymptotic regime of  $p_-\lesssim 0.05$ that was identified in  Fig.~\ref{fig:sEstimxx}(b), albeit for a different model.      Setting   $\langle T_Y\rangle \approx \delta/v$,  with $\delta = 8{\rm nm}$ the step size of the molecular motor, and  considering room temperatures   $\mathsf{T}_{\rm env} = 298 \times 1.38 \times 10^{-23} {\rm J}$, we obtain 
\begin{equation}
\eta \leq  \eta_{\rm FPR} =  \frac{1}{1+\frac{298  \times 1.38 \times  0.01}{24   } \times \left|\ln  0.02\right|} \approx  0.6.   
\end{equation} 
As this  simple back-of-the-envelope calculation illustrates, we can  estimate dissipation in nonequilibrium processes with  $\hat{s}_{\rm FPR}$.    In the following sections, we use a more detailed analysis that relies on a model for Kinesin-1 dynamics to determine the accuracy of this estimate.

%which should be compared with an  efficiency of  $0.47$ estimated in Ref.~\cite{hwang2018energetic, hwang2019correction} using a model,   and the bound from the TUR uncertainty relations of $0.7$.      Hence, in this simple back on the envelope , both FPR and TUR provide similar estimates on the efficiency of Kinesin-1.       As we will see in the next sections, this is because the thresholds are small, i.e. $\ell_-=\ell_+ = 8 {\rm nm}$.

\subsection{Model for  molecular motors with two motor heads}\label{sec:52}   
In the previous section, we have used the FPR to estimate the rate of dissipation of a Kinesin-1 motor.      In this section, we  consider a molecular motor model for Kinesin-1 to determine the accuracy of this estimate.   

We describe the dynamics of Kinesin-1 with a model for two-headed molecular motors.   This model is a  Markov jump  process  that determines the  transitions between the chemical states of the motor heads, and the model includes also one mechanical step.         Each motor head can take three different chemical states, namely, the ADP-bound (${\rm D}$),  ATP-bound (${\rm T}$), and nucleotide-free APO states ($\phi$), leading to nine possible states.     However,  since the kinetics of the  two heads are coordinated, see Refs.~\cite{PhysRevLett.98.258102, hwang2018energetic}, a six state model  with the states 
\begin{equation}
X(t) \in \left\{ ({\rm D}:\phi), ({\rm D}:{\rm T}), (\phi:{\rm T}),  (\phi:{\rm D}),({\rm T}:D),  ({\rm T}:\phi) \right\}
\end{equation}
suffices to describe  the dynamics of  two-headed molecular motors.    For convenience, we also number the states from $1$ to $6$, as shown in Fig.~\ref{fig:examples}(b).    

 In Fig.~\ref{fig:examples}(b) we show all the possible chemical/mechanical transitions in the  six-state  model  for two-headed molecular motors.   As argued in Ref.~\cite{PhysRevLett.98.258102}, the mechanical step,  corresponding with the interchange of  the front and rear head, takes place during  the transition from $({\rm D}:{\rm T})$ to $({\rm T}:{\rm D})$, and therefore the position of the motor equals the current
 \begin{equation}
 Y(t) := J_{2\rightarrow 5}(t),
 \end{equation}
 where for convenience we have set $Y(0)=0$.  

The model contains three cycles: a forward cycle at  rate $\overline{j}_{\rm f}$ corresponding to an  ATP-hydrolysis-induced forward step,    a backward cycle at rate $\overline{j}_{\rm b}$ corresponding to an ATP-hydrolysis-induced backward step, and a third cycle with rate $\overline{j}_0$ that hydrolyses two ATP molecules but does not change the position of the motor.   The total rate of dissipation is  
\begin{equation}
\dot{s} = \overline{j}_{\rm f} \frac{a_{\rm f}}{\mathsf{T}_{\rm env}}   + \overline{j}_{\rm b} \frac{a_{\rm b}}{\mathsf{T}_{\rm env}}  +\overline{j_{0}} \frac{a_0}{\mathsf{T}_{\rm env}}
\end{equation}
where 
\begin{equation}
\frac{a_{\rm f}}{\mathsf{T}_{\rm env}}   =  \ln \frac{k_{1\rightarrow 2}k_{2\rightarrow 5}k_{5\rightarrow 6}k_{6\rightarrow 1}}{k_{1\rightarrow 6}k_{6\rightarrow 5}k_{5\rightarrow 2}k_{2\rightarrow 1}} = \frac{\Delta \mu}{\mathsf{T}_{\rm env}}  - \frac{f_{\rm mech} \delta }{\mathsf{T}_{\rm env}} ,
\end{equation}  
\begin{equation}
\frac{a_{\rm b}}{\mathsf{T}_{\rm env}}   =  \ln \frac{k_{2\rightarrow 3}k_{3\rightarrow 4}k_{4\rightarrow 5}k_{5\rightarrow 2}}{k_{2\rightarrow 5}k_{5\rightarrow 4}k_{4\rightarrow 3}k_{3\rightarrow 2}} =\frac{ \Delta \mu }{\mathsf{T}_{\rm env}}  + \frac{ f_{\rm mech} \delta }{\mathsf{T}_{\rm env}} ,
\end{equation} 
are the affinities of the forward and backward cycles, and   
\begin{equation}
a_0  = a_{\rm f} + a_{\rm b}
\end{equation}
is the affinity of the third cycle.      Here 
\begin{equation} 
 \Delta \mu = \mathsf{T}_{\rm env}   \ln  \left(K_{\rm eq} \frac{[{\rm ATP}]}{[{\rm ADP}][{\rm P}] }\right) \label{eq:deltamux}
 \end{equation}
 is  the supplied chemical energy due to the hydrolysis of ATP into ADP and P,  and $f_{\rm mech}\delta$ is  the mechanical work  done by the system on an external cargo  during one molecular motor step of length $\delta$.     In Eq.~(\ref{eq:deltamux}), [ATP], [ADP], and [P], are  the concentrations  in the surrounding chemostat of ATP, ADP, and P, respectively,   and $K_{\rm eq}$ is the equilibrium constant of the hydrolysis reaction.

Following \cite{PhysRevLett.98.258102, hwang2018energetic},  we use for the rates of the chemomechanical transitions between states $2$ and $5$ the formulae
\begin{equation}
k_{2\rightarrow 5}(f_{\rm mech} ) =  k_{2\rightarrow 5}(0) e^{-\theta\frac{f_{\rm mech} \delta}{\mathsf{T}_{\rm env}}}  \label{eq:param1}
 \end{equation}
 and 
 \begin{equation}
 k_{5\rightarrow 2}(f_{\rm mech} ) =  k_{5\rightarrow 2}(0) e^{(1-\theta) \frac{f_{\rm mech} \delta}{\mathsf{T}_{\rm env}}}.\label{eq:param2}
  \end{equation} 
  For the other rates we set 
  \begin{equation}
  k_{i\rightarrow j}(f_{\rm mech}) = \frac{2k_{i\rightarrow j}(0)}{1+e^{\chi_{ij}\frac{f_{\rm mech}\delta}{\mathsf{T}_{\rm env}}}}\label{eq:param3}
  \end{equation}
  with $\chi_{ij} = \chi_{ji}$.     The concentration $[{\rm ATP}]$ determines the rates   
  \begin{equation}
k_{1\rightarrow 2}(0) = k^{\rm bi}_{1\rightarrow 2}[{\rm ATP}]
  \end{equation}
  and 
   \begin{equation}
k_{4\rightarrow 5}(0) = k^{\rm bi}_{4\rightarrow 5}[{\rm ATP}], 
  \end{equation}    
 and the explicit dependence on the  concentrations $[{\rm ADP}]$ and $[{\rm P}]$ is not considered.    Due to the equivalence of the two cycles, we also set $k_{3\rightarrow 2}(0) = k_{6\rightarrow 5}(0) $, $k_{2\rightarrow 3}(0) =k_{5\rightarrow 6}(0) $, $k_{3\rightarrow 4}(0)  = k_{6\rightarrow 1}(0) $, $k_{4\rightarrow 3}(0)  = k_{1\rightarrow 6}(0) $, $k_{4\rightarrow 5}(0)  = k_{1\rightarrow 2}(0) $, $\chi_{23}=\chi_{56}$, $\chi_{34}=\chi_{61}$, and $\chi_{45}=\chi_{12}$.    Moreover, we set
 \begin{equation}
    k_{5\rightarrow 2}(0) = k_{2\rightarrow 5}(0) \sqrt{\frac{k_{5\rightarrow 4}(0)}{k_{2\rightarrow 1}(0)}}, \label{eq:K52}
\end{equation}
such that the six-state model satisfies detailed balance when 
\begin{equation}
f_{\rm mech}=0  \quad {\rm and} \quad  [{\rm ATP}] =  \frac{k_{1\rightarrow 6}(0)}{k_{6\rightarrow 1}(0)} \frac{k_{6\rightarrow 5}(0)}{k_{5\rightarrow 6}(0)} \frac{\sqrt{k_{5\rightarrow 4}(0)k_{2\rightarrow 1}(0)} }{k^{\rm bi}_{4\rightarrow 5}}.\label{eq:ATP}  
\end{equation}
For these values of the control parameters, there is neither a chemical driving force  ($\Delta \mu=0$) nor a mechanical force ($f_{\rm mech}=0$). 
  
The remaining parameters $k_{i\rightarrow j}(0)$, $\theta$,  $\chi_{ij}$, $k^{\rm bi}_{1\rightarrow 2}$, and $k^{\rm bi}_{4\rightarrow 5}$ are motor specific, and for Kinesin-1 they have been obtained in Refs.~\cite{hwang2018energetic, hwang2019correction} from fitting single molecule motility data to the  six-state model.   Here, we will use these fitted parameters for Kinesin-1, and for  convenience we summarise the exact numerical values in \ref{app:C}.    

The concentration $[{\rm ATP}]$ and the mechanical load $f_{\rm mech}\delta/\mathsf{T}_{\rm env}$ are  external, control parameters that can be modified by the experimenter.

\begin{figure}[t!]
\centering
\includegraphics[width=0.65\textwidth]{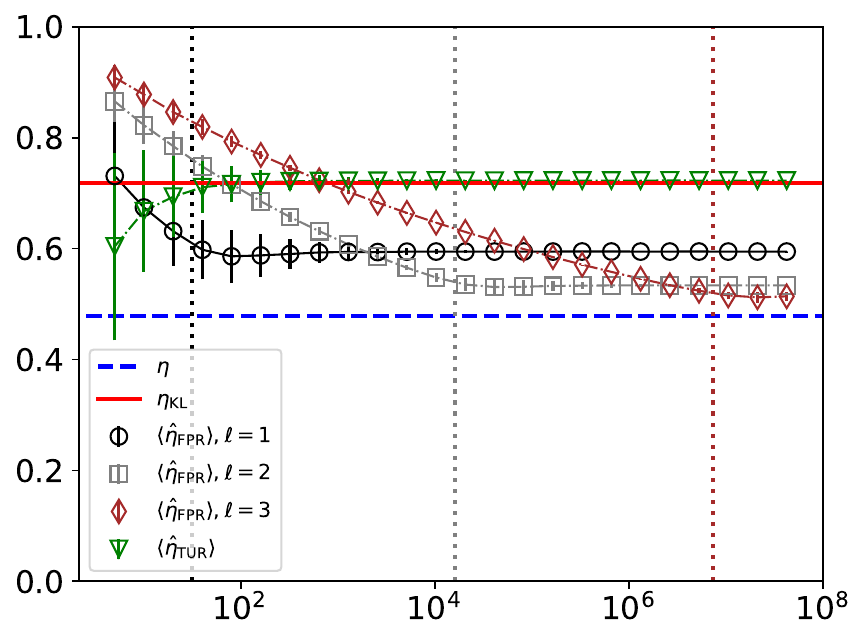}
  \put(-150,-10){\Large$n_{\rm s}$} 
\caption{{\it Simulation results for  estimated molecular motor efficiencies.} The estimated efficiencies $\langle \hat{\eta}_{\rm FPR}\rangle$, based on first-passage processes for the position $Y(t)$ in the molecular motor model  described in Sec.~\ref{sec:52}, as a function of the number of realisations $n_{\rm s}$ for three different values of the thresholds $\ell_-=\ell_+=\ell$ (as indicated by the legend), $[\rm ATP]=10\mu {\rm M}$, and $f_{\rm mech} = 3{\rm pN}$.   For reference, the efficiency $\eta = 0.478$ is shown with the horizontal, blue dashed line; the estimate $\eta_{\rm KL}$ based on the Kullback-Leibler divergence in the limit of $n_{\rm s}\gg1$ is shown by the horizontal, red, solid line; and the  estimates  $\langle \hat{\eta}_{\rm TUR}\rangle$ based on the TUR evaluated at $t=1 {\rm s}$ are shown by the green triangles.       Each marker is an average over $1000$ or $100$ (for large values of $n_{\rm s}$) realisations and the error bars denote the standard deviations of $\hat{\eta}_{\rm FPR}$ and $\hat{\eta}_{\rm TUR}$.    The vertical, dotted lines denote the values of $n_{\rm s}=1/p_-$.  } \label{fig:sEstim3}
\end{figure}

  \subsection{Estimating the efficiency of a molecular motor   based on the measurement of its  position}
We estimate the efficiency $\eta$  from the measurement of the position $Y$ of the molecular motor, which according to Eq.~(\ref{eq:sEff})  is equivalent to an entropy production estimation problem.     We consider  the case of $[{\rm ATP}] = 10 \mu {\rm M}$ and $f_{\rm mech}=3 {\rm pN}$, corresponding with the calculation in Sec.~\ref{sec:52x} based on experimental data. 

We define two estimators     $\hat{\eta}_{\rm FPR}$,  and $\hat{\eta}_{\rm TUR}$ for the efficiency $\eta$  that are  equivalent to the estimators $\mathcal{F}_{\rm FPR}$ and $\mathcal{F}_{\rm TUR}$ for the rate of dissipation, i.e., they are  estimates for the efficiency when a finite number of trajectories are available.      For estimation based on   first-passage processes, we consider 
 \begin{equation}
 \hat{\eta}_{\rm FPR} := \left(1+   \frac{\mathsf{T}_{\rm env}}{f_{\rm mech} v} \frac{\ell_+}{\ell_-}\frac{ \left|\ln  \hat{p}_- \right|}{\langle \hat{T}_Y\rangle}\right)^{-1}, 
 \end{equation}
 where $\hat{p}_-$  and $\langle\hat{T}_Y\rangle$ are the estimates of $p_-$ and $\langle T_Y\rangle$ based on a finite number of realisations of the process,  as in Eqs.~(\ref{eq:estimatedTJ}) and (\ref{eq:estimatedpM}); note that  we assume that   $v = \delta\langle Y(t)\rangle/t$ is known, although we could also take an estimated value for this quantity leading to almost identical results.     To estimate the motor efficiency based on the TUR, we  use 
   \begin{equation}
\hat{\eta}_{\rm TUR} := \left(1+\frac{ \mathsf{T}_{\rm env} }{f_{\rm mech} \delta\: }\frac{2 \hat{Y}(t)}{\hat{\sigma}^2_{Y(t)}}\right)^{-1},
\end{equation} 
where $\hat{Y}(t)  = n_{\rm s}^{-1}\sum^{n_{\rm s}}_{i=1}Y^{(i)}(t)$ and $\hat{\sigma}^2_{Y(t)} =n^{-1}_{\rm s} \sum^{n_{\rm s}}_{i=1}\left(Y^{(i)}(t)\right)^2   - n^{-2}_{\rm s} \left(\sum^{n_{\rm s}}_{i=1}Y^{(i)}(t)\right)^2$.

 Figure \ref{fig:sEstim3}  plots  $\langle \hat{\eta}_{\rm FPR}\rangle$ as a function of $n_{\rm s}$ for three values of $\ell=\ell_-=\ell_+$, namely, $\ell=1$, $\ell=2$, and $\ell=3$, and compares the estimate $\langle \hat{\eta}_{\rm FPR}\rangle$   with the efficiency $\eta = 0.478$ and the estimate    $\langle \hat{\eta}_{\rm TUR} \rangle$ based on the TUR. 
 
 Just as for the $\mathcal{F}_{\rm FPR}$  in Fig.~\ref{fig:samples},  we can differentiate two regimes: (i) for $n_{\rm s}<1/p_-$, the estimator  $\langle\hat{\eta}_{\rm FPR}\rangle$ decreases logarithmically; and (ii) for $n_{\rm s}>1/p_-$, the estimator  $\langle\hat{\eta}_{\rm FPR}\rangle$ has reached its asymptotic value.      The values $n_{\rm s} = 1/p_-$ that separate the two regimes are indicated for different values of $\ell$ by the vertical dotted lines.    
 
 For $\ell=1$, $\langle \hat{\eta}_{\rm FPR} \rangle$ converges to its  asymptotic value $0.6$, which is similar to the value we estimated in Sec.~\ref{sec:52x} based on experimental data.   The simulation results in Fig.~\ref{fig:sEstim3}  show that this estimate can be improved by considering larger values of  $\ell$, but this comes at a significant expense in the number of realisations $n_{\rm s}$ required.    Nevertheless, the estimate  $\langle \hat{\eta}_{\rm FPR}\rangle$  at small thresholds $\ell=1$ is better than the estimate  $\langle \hat{\eta}_{\rm TUR}\rangle$, indicating again that small thresholds are sufficient to obtain accurate estimates of dissipation with FPRs.       
 
Interestingly, Fig.~\ref{fig:sEstim3}  shows that the estimates  $\langle \hat{\eta}_{\rm FPR} \rangle$ are less biased than the estimate $\eta_{\rm KL}$, given by Eq.~(\ref{eq:KLEeta}), based on the Kullback-Leibler divergence.     Note that in the present example  $Y$ is a nonMarkovian current.  Since $\eta_{\rm KL}$  does not account for nonMarkovian effects in the current, we can attribute the smaller bias in the FPR due to the nonMarkovian statistics of $Y$ that are accounted for by the    FPR.

\section{Discussion}   \label{sec:6} 
We have studied the problem of inferring   the rate of entropy production $\dot{s}$  from the  measurement of  repeated realisations of a first-passage process  of a single current.   To this aim, we have used the estimator $\mathcal{F}_{\rm FPR}$ [Eq.~(\ref{eq:FFPR})]  that converges in the limit of an infinite number of realisations of the process to the  first-passage ratio $\hat{s}_{\rm FPR}$ [Eq.~(\ref{eq:FPR})].        We have compared this estimator with  $\mathcal{F}_{\rm TUR}$, an estimator based on the TUR, and with $\mathcal{F}_{\rm KL}$, an estimator based on the Kullback-Leibler divergence.  We discuss now  the strengths and weaknesses of these different estimators, and we also provide an outlook on future challenges. 

Comparing  $\mathcal{F}_{\rm FPR}$ with $\mathcal{F}_{\rm TUR}$, we conclude that far from thermal equilibrium the FPR is a better estimator of dissipation.   Indeed, as shown in Fig.~\ref{fig:sEstim2}, far from thermal equilibrium the  FPR captures a finite fraction of  $\dot{s}$, while the TUR captures a negligible fraction of $\dot{s}$; note that in the literature modifications of the TUR have been derived  that have a smaller bias than $\hat{s}_{\rm TUR}$~\cite{shiraishi2021optimal, PhysRevX.11.041061}, but     these modified  TURs rely on measurements of both currents and  time-integrated empirical measures, and hence require more information about the process $X$ than the trajectories of a single current.    The estimation of dissipation with $\hat{s}_{\rm FPR}$ does require a number $n_{\rm s}\gg 1/p_-$ of realisations of the nonequilibrium process,  which can be large for processes far from thermal equilibrium; in particular, we have shown that the number of realisations $n_{\rm s}$ increases exponentially as a function of $\Delta s$, the minimal jump size of $S$.     Nevertheless, since  far from thermal equilibrium $\mathcal{F}_{\rm TUR}/\dot{s}\approx 0$, it holds that the bias of  the FPR is smaller than the bias of the TUR, even for values   of $n_{\rm s}$ that are smaller than $1/p_-$.

Comparing $\mathcal{F}_{\rm FPR}$ with $\mathcal{F}_{\rm KL}$, we conclude that for generic Markov jump processes the bias in    $\mathcal{F}_{\rm FPR}$  is smaller than the bias in  $\mathcal{F}_{\rm KL}$,  as the FPR captures nonMarkovian effects in the current (see Fig.~\ref{fig:sEstim3}).    An exception is  when the current $J$ is Markovian, as was the case for the random walk  process discussed in Sec.~\ref{sec:4}.   Note that it is possible to include memory effects in the Kullback-Leibler divergence, e.g., by assuming the current is a semi-Markov process \cite{martinez2019inferring}, but in general   this  still leads to a biased estimator of dissipation.

Taken together, the  three estimators $\mathcal{F}_{\rm FPR}$, $\mathcal{F}_{\rm TUR}$, and $\mathcal{F}_{\rm KL}$ are valid estimators of dissipation, as they can be used to estimate dissipation in different circumstances.   Near equilibrium and with limited data available, the TUR is a good estimator of dissipation; for Markovian currents the Kullback-Leibler divergence is the best estimator; and for processes far from equilibrium with nonMarkovian currents the FPR is the estimator with the least bias.    

In the present paper, we have considered  entropy production estimation based on the measurement of a single current.      Since all estimators are bounded from above by the rate $\dot{s}$ of entropy production, we can reduce the bias in the estimators by measuring multiple currents, and maximising the estimator over linear combinations of these currents.   In this regard, the FPR has the appealing property that the maximum of this variational problem is attained by a current that is proportional to the entropy production, and hence the affinities of the process can be inferred from a variational principle.

%Antother interesting research avenue is to infer dissipation rates of  feasilibty of entropy production estimation in biological cells.   With current advances in superresolution microscopy  it is possible to live-cell image the trajectories of proteins at a resolution of $50{\rm nm}$ [....].   It will bre important to determine whether this resolution is precise enough for dissipation.   

 The purpose of the present paper was to highlight the potential of a first-passage approach to the entropy-production estimation problem.  To this aim, we have performed a comparative study using two simple examples of nonequilibrium processes.    However,   to better understand estimation theory for entropy production rates,  we think it  will be important to take the present study one step further and make  an extensive comparison of the different methods for the estimation of entropy production used in the literature, e.g., by using a larger pool of estimators for entropy production  and a larger pool of models representing nonequilibrium systems.   In this regard, one should  be careful to make a fair comparison of the different methods based on the  knowledge they require about the nonequilibrium process and the amount of data used.  Also, we hope the present manuscript highlights the importance of   developing further  the  mathematics underlying the estimator $\hat{s}_{\rm FPR}$ through extensions of the   results in Ref.~\cite{neri2021universal}.   

We end the paper with the discussion of an interesting, potential application in  cell biology.    Proteins in living cells produce entropy, as they are involved in chemical reactions that operate far from chemical equilibrium.    However, due to the diffraction limit, it is  challenging to measure the trajectories of individual proteins in living cells at high enough spatial and temporal resolution.   Nevertheless, in recent years significant advances have been made in super-resolution live cell imaging, see e.g.~Refs.~\cite{fernandez2008fluorescent, lelek2021single, ranjan2021super}.  If the resolution of these imaging methods can be further improved, then it will  be  possible to   infer entropy production rates of individual proteins in living cells  from the measurements of their trajectories.

   \appendix

\section{Splitting probabilities of $S$}\label{App:a}
To derive an exact expression for the splitting probabilities $p_-$
 and $p_+$ of $S(t)$, we use the integral fluctuation relation at stopping times \cite{neri2019integral}
\begin{equation}
\langle e^{-S(T_J)} \rangle=1,
\end{equation}
where $T_J$ is a first-passage time of a current $J$ with two thresholds, as defined in Eq.~(\ref{eq:FPTime}).   The derivations below follow the  Refs.~\cite{neri2017statistics, neri2019integral}. 
 
 Since the interval $(-\ell_-,\ell_+)$ is finite, 
\begin{equation}
\mathbb{P}\left(T_J<\infty\right)=1, 
\end{equation}
and therefore
\begin{equation}
p_- + p_+ = 1 \label{eq:pSum}
\end{equation}
and 
\begin{equation}
p_- \langle e^{-S(T_J)} | S(T_J)\leq -\ell_-\rangle + p_+ \langle e^{-S(T_J)} | S(T_J)\geq \ell_+\rangle  =1, \label{eq:pSum2}
\end{equation}
where $\langle \cdot|\cdot\rangle$ are conditional expectations.  
Equations (\ref{eq:pSum}) and (\ref{eq:pSum2}) imply that 
\begin{equation}
p_- =  \frac{1-\langle e^{-S(T_J)}|S(T_J)\geq \ell_+\rangle }{\langle e^{-S(T_J)} | S(T_J)\leq -\ell_-\rangle   -  \langle e^{-S(T_J)} | S(T_J)\geq \ell_+\rangle} .  \label{eq:integral2}
\end{equation}

Hence, when the stochastic entropy production $S(t)$ has continuous trajectories --- for Markov jump processes this is the near-equilibrium limit --- Eq.~(\ref{eq:integral2}) implies 
\begin{equation}
p_- =  \frac{1-e^{-\ell_+} }{e^{\ell_-}  -  e^{-\ell_+}} .
\end{equation}

On the other hand,  when the stochastic entropy production $S(t)$ has jumps ---   for Markov jump processes this is the general  case ---  this gives
\begin{equation}
p_- \leq  \frac{1}{e^{\ell_-}  -  e^{-\ell_+}} . \label{eq:boundpM}
\end{equation}
In addition, when the depth $\ell_-$ of the negative thresholds is small, the bound (\ref{eq:boundpM}) can be improved.   Particularly, defining 
\begin{equation}
\Delta s  = {\rm inf}\left\{|S(t)-\lim_{\epsilon\rightarrow 0^+}S(t-\epsilon)|:t\geq 0\right\},
\end{equation}
the infimum value of the jump size in $S(t)$ (the infimum applies only to the jump times), we obtain  
\begin{equation}
p_- \leq  \frac{1}{e^{ {\rm max}\left\{\ell_-,\Delta s\right\}}  -  e^{-\ell_+}} .
\end{equation} 
Hence, for  processes governed far from thermal equilibrium so that $\Delta s\gg 1$, we obtain 
\begin{equation}
p_- \leq e^{-\Delta s}.
\end{equation}

\section{Recursion equations for $\langle T_J\rangle$ and $p_-$ for a  random walker on a two-dimensional lattice}  \label{App:pM}
 We derive a set of recursion equations that determine $\langle T_J\rangle$ and $p_-$ for currents $J$, as given by Eq.~(\ref{eq:JM}),  in   random walkers  on a two-dimensional lattice.
 
\subsection{Recursion equations for $p_-$}

Let $p_-(x_1,x_2)$  be the probability for $X=(X_1,X_2)$ to reach  the region
\begin{equation}
(1-\Delta)X_1 + (1+\Delta)X_2  \leq -\ell_-
\end{equation}  
before reaching 
\begin{equation}
(1-\Delta)X_1  + (1+\Delta)X_2  \geq \ell_+,
\end{equation}  
when the initial values $X_1(0) = x_1$ and $X_2(0)=x_2$.

Let 
\begin{equation}
\mathcal{D}_- =  \left\{(x,y)\in\mathbb{R}^2:  (1-\Delta)x + (1+\Delta)y \leq -\ell_-\right\}
\end{equation}
and 
\begin{equation}
\mathcal{D}_+=  \left\{(x,y)\in\mathbb{R}^2:  (1-\Delta)x + (1+\Delta)y \geq  \ell_+\right\}.
\end{equation}

It  holds for all $(x,y)\notin \mathcal{D}_-\cup \mathcal{D}_+$   that \cite{doyle1984random}
  \begin{eqnarray}
\fl   p_-(x,y) =  \frac{k^-_1}{ k_{\rm tot}} p_-(x-1,y) + \frac{k^+_1}{k_{\rm tot}}  p_-(x+1,y)   
  +\frac{k^-_2}{k_{\rm tot}} p_-(x,y-1) + \frac{k^+_2}{k_{\rm tot}}  p_-(x,y+1),  \nonumber\\   \label{eq:pL}
  \end{eqnarray}
  where $k_{\rm tot} =  k^+_1+k^-_1+k^+_2+k^-_2$, and with  boundary conditions 
    \begin{equation}
    p_-(x,y) = 1  \quad {\rm if} \quad   (x,y) \in \mathcal{D}_-\label{eq:pL1}
      \end{equation}
      and 
    \begin{equation}
    p_-(x,y) = 0  \quad {\rm if} \quad   (x,y) \in \mathcal{D}_+.\label{eq:p2}
          \end{equation}
      
The Eqs.~(\ref{eq:pL}-\ref{eq:p2}) admit a unique solution $p_-(x,y)$ \cite{doyle1984random}, which can be found by solving the iterative equations 
     \begin{eqnarray}
\fl   p_-(x,y;n) =  \frac{k^-_1}{ k_{\rm tot}} p_-(x-1,y;n-1) + \frac{k^+_1}{k_{\rm tot}}  p_-(x+1,y;n-1)   \nonumber\\ 
  +\frac{k^-_2}{k_{\rm tot}} p_-(x,y-1;n-1) + \frac{k^+_2}{k_{\rm tot}}  p_-(x,y+1;n-1), \label{eq:pLt}
  \end{eqnarray} 
   for  $(x,y)\in \mathbb{R}^2\setminus \mathcal{D}^-\cup\mathcal{D}^+$ and $n\in\mathbb{N}$, and with 
  boundary  conditions 
     \begin{equation}
    p_-(x,y;n) = 1 \quad {\rm if} \quad   (x,y) \in \mathcal{D}_-
      \end{equation}
      and 
    \begin{equation}
    p_-(x,y;n) = 0 \quad {\rm if} \quad   (x,y) \in \mathcal{D}_+, \label{eq:p2t}
      \end{equation}  
 also       for  $n\in\mathbb{N}$.
      In the limit of large $n$,  the stationary solution to the Eqs.~(\ref{eq:pLt}-\ref{eq:p2t})    converges to the solution of (\ref{eq:pL}-\ref{eq:p2}).   
      
\subsection{Recursion equations for $\langle T_J\rangle$}

The mean first-passage time $t(x,y) = \langle T_J|X_1(0)=x,X_2(0)=y\rangle$ solves for all   $(x,y)\notin \mathcal{D}_-\cup \mathcal{D}_+$ the equations \cite{doyle1984random} 
  \begin{eqnarray}
\fl   t(x,y) =  \frac{k^-_1}{ k_{\rm tot}} t(x-1,y)+ \frac{k^+_1}{k_{\rm tot}}  t(x+1,y)  
  +\frac{k^-_2}{k_{\rm tot}} t(x,y-1) + \frac{k^+_2}{k_{\rm tot}}  t(x,y+1) + \frac{1}{k_{\rm tot}}   
  \end{eqnarray}
  and 
  with boundary conditions 
    \begin{equation}
    t(x,y) = 0  \quad {\rm if} \quad   (x,y) \in \mathcal{D}_-
      \end{equation}
      and 
    \begin{equation}
    t(x,y) = 0  \quad {\rm if} \quad   (x,y) \in \mathcal{D}_+.
      \end{equation} 
      
      Just as before, we can solve these equations iteratively by introducing a time-index 
        \begin{eqnarray}
\fl   t(x,y;n) =  \frac{k^-_1}{ k_{\rm tot}} t(x-1,y;n-1)+ \frac{k^+_1}{k_{\rm tot}}  t(x+1,y;n-1)   \nonumber\\ 
  +\frac{k^-_2}{k_{\rm tot}} t(x,y-1;n-1) + \frac{k^+_2}{k_{\rm tot}}  t(x,y+1;n-1) + \frac{1}{k_{\rm tot}} ,
  \end{eqnarray}
     for  $(x,y)\in \mathbb{R}^2\setminus \mathcal{D}^-\cup\mathcal{D}^+$ and  $n\in\mathbb{N}$, and with
    \begin{equation}
    t(x,y;n) = 0  \quad {\rm if} \quad   (x,y) \in \mathcal{D}_-
      \end{equation}
      and 
    \begin{equation}
    t(x,y;n) = 0  \quad {\rm if} \quad   (x,y) \in \mathcal{D}_+,
      \end{equation} 
 also     for all $n\in\mathbb{N}$.

\section{Parameters used for the six-state model of Kinesin-1}\label{app:C}
We specify the parameters that we use in Sec.~\ref{sec:5} for the  six-state model for two-headed molecular motors, as visualised in Fig.~\ref{fig:examples}.      

The rates are parametrised as in  Eqs.~(\ref{eq:param1}), (\ref{eq:param2}), and (\ref{eq:param3}), and we  thus need to specify the $k_{i\rightarrow j}(0)$, $\theta$, and $\chi_{ij}$.       All parameters, except for $k_{5\rightarrow 2}(0)$, are set identical as those in Ref.~\cite{hwang2018energetic}, which 
were obtained by fitting numerical results to single molecule motility data.    In particular we set $\theta = 0.61$, $\chi_{12} = 0.15$, $\chi_{56} = 0.0015$, $\chi_{61} = 0.11$, $k_{2\rightarrow 1}(0) = 4200 \: s^{-1}$, $k_{2\rightarrow5}(0) = 1.6 \times 10^6 \:  s^{-1}$,$k_{5\rightarrow6}(0)=190 \:  s^{-1}$, $k_{6\rightarrow5}(0)=10 \:  s^{-1}$, $k_{6\rightarrow1}(0) = 250 \:  s^{-1}$, $k_{1\rightarrow6}(0) = 230\:   s^{-1}$, $k_{5\rightarrow4}(0)= 2.1 \times 10^{-9} \:  s^{-1}$, and we set  $k_{1\rightarrow 2}(0) = k^{\rm bi}_{1\rightarrow 2}(0)[{\rm ATP}]$ with $k^{\rm bi}_{1\rightarrow 2}(0) = 2.8\:   \mu M^{-1}s^{-1}$.    
Note that due to the equivalence of the two cycles, we have set $k_{3\rightarrow 2} = k_{6\rightarrow 5}$, $k_{2\rightarrow 3}=k_{5\rightarrow 6}$, $k_{3\rightarrow 4} = k_{6\rightarrow 1}$, $k_{4\rightarrow 3} = k_{1\rightarrow 6}$, $k_{4\rightarrow 5} = k^{\rm bi}_{4\rightarrow 5}[{\rm ATP}] = k_{1\rightarrow 2} =k^{\rm bi}_{1\rightarrow 2}[{\rm ATP}]  $, $\chi_{23}=\chi_{56}$, $\chi_{34}=\chi_{61}$, and $\chi_{45}=\chi_{12}$.

For the rate $k_{5\rightarrow 2}(0)$ we have used Eq.~(\ref{eq:K52}) 
instead of the value
\begin{equation}
    k_{5\rightarrow 2}(0) = 1.1 s^{-1}
\end{equation}
used in Ref.~\cite{hwang2018energetic}.       The reason we prefer to use Eq.~(\ref{eq:K52})  is because then the model satisfies detailed balance when the control parameters are given by Eq.~(\ref{eq:ATP}).   Moreover,  since 
\begin{equation}
 k_{2\rightarrow 5}(0) \sqrt{\frac{k_{5\rightarrow 4}(0)}{k_{2\rightarrow 1}(0)}} \approx  1.13 {\rm s}^{-1}
\end{equation}
this will make no discernible difference in the  model fits.

At equilibrium, i.e. for the control parameters given by Eq.~(\ref{eq:ATP}), the stationary state is 
\begin{eqnarray}
\fl p_{\rm eq}(1)=\frac{1}{\mathcal{N}}, \quad  p_{\rm eq}(2) =\frac{1}{\mathcal{N}}\sqrt{\frac{k_{5\rightarrow 4}(0)}{k_{2\rightarrow 1}(0)}}\frac{k_{4\rightarrow 3}(0) }{k_{3\rightarrow 4}(0)}  \frac{k_{3\rightarrow 2}(0)}{k_{2\rightarrow 3}(0)},   \quad p_{\rm eq}(3) =\frac{1}{\mathcal{N}} \sqrt{\frac{k_{5\rightarrow 4}(0)}{k_{2\rightarrow 1}(0)}}\frac{k_{4\rightarrow 3}(0) }{k_{3\rightarrow 4}(0)},  \nonumber \\ 
\fl p_{\rm eq}(4) = \frac{1}{\mathcal{N}}\sqrt{\frac{k_{5\rightarrow 4}(0)}{k_{2\rightarrow 1}(0)}},  \quad p_{\rm eq}(5) = \frac{1}{\mathcal{N}}\frac{k_{4\rightarrow 3}(0) }{k_{3\rightarrow 4}(0)}  \frac{k_{3\rightarrow 2}(0) }{k_{2\rightarrow 3}(0)},  \quad  p_{\rm eq}(6) = \frac{1}{\mathcal{N}}\frac{k_{4\rightarrow 3}(0) }{k_{3\rightarrow 4}(0) },
\end{eqnarray}
where the normalisation constant is 
\begin{eqnarray}
\fl \mathcal{N} = 1 + \sqrt{\frac{k_{5\rightarrow 4}(0)}{k_{2\rightarrow 1}(0)}}\frac{k_{4\rightarrow 3}(0) }{k_{3\rightarrow 4}(0)}  \frac{k_{3\rightarrow 2}(0) }{k_{2\rightarrow 3}(0)}  +  \sqrt{\frac{k_{5\rightarrow 4}(0)}{k_{2\rightarrow 1}(0)}}\frac{k_{4\rightarrow 3}(0) }{k_{3\rightarrow 4}(0)}  
\nonumber \\ 
+\sqrt{\frac{k_{5\rightarrow 4}(0)}{k_{2\rightarrow 1}(0)}}  + \frac{k_{4\rightarrow 3}(0) }{k_{3\rightarrow 4}(0)}  \frac{k_{3\rightarrow 2}(0) }{k_{2\rightarrow 3}(0)} + \frac{k_{4\rightarrow 3}(0) }{k_{3\rightarrow 4}(0) }.
\end{eqnarray}

\section*{References}
\bibliographystyle{ieeetr} % plain, alpha, ieeetr
\bibliography{biblio}

\end{document}